\documentclass[preprint,aps]{revtex4-2}
\usepackage{mathrsfs}
\usepackage{graphicx}
\usepackage{multirow}
\usepackage{makecell}
\usepackage{booktabs}
\usepackage{color}
\usepackage{bm}
\usepackage{amsmath}
\usepackage{hyperref}
\hypersetup{
    colorlinks,
    citecolor=black,
    filecolor=black,
    linkcolor=black,
    urlcolor=black
}
\begin{document}

\title{Floquet dressing of bulk metallic bands in 2H-NbSe$_2$}

\author{Cong Li$^{1,2,\sharp,*}$, Qirui Cui$^{2,\sharp,*}$, Maciej Dendzik$^{2}$, Magnus H. Berntsen$^{2}$, Wanyu Chen$^{2}$, Zhilin Li$^{1}$, Anna Delin$^{2}$, Wentao Zhang$^{1,*}$, Oscar Tjernberg$^{2,*}$
}

\affiliation{
\\$^{1}$Beijing National Laboratory for Condensed Matter Physics, Institute of Physics, Chinese Academy of Sciences, Beijing 100190, China
\\$^{2}$Department of Applied Physics, KTH Royal Institute of Technology, Stockholm 11419, Sweden
\\$^{\sharp}$These authors contributed equally to this work.
\\$^{*}$Corresponding authors: licong@iphy.ac.cn, qiruic@kth.se, wentaozhang@iphy.ac.cn, oscar@kth.se
}


\maketitle


\begin{center}
{\bf Abstract}
\end{center}

{\bf Floquet dressing offers a route to controlling electronic states through periodic optical driving~\cite{TOka_ARCM2019_SKitamura,MSRudner_NRP2020_NHLindner}. Its realization in solids has centred primarily on topological-insulator surface states, semiconductors and two-dimensional Dirac semimetals~\cite{YHWang_Science2013_NGedik,SZhou_Nature2023_SZhou,DChoi_NP2025_NGedik,MMerboldt_NP2025_SMathias}. Extending this control to bulk metals would enable manipulation of the itinerant states underlying metallic transport and collective order. Yet strong screening and rapid scattering have long cast doubt on its feasibility in bulk metals, while photoelectron dressing complicates its identification~\cite{STPark_PRA2014_STPark}. Here we identify Floquet dressing of the overlapping, bulk-derived metallic bands of 2H-NbSe$_2$ using time- and angle-resolved photoemission spectroscopy. Photon-shifted replicas in a sparse unoccupied-state window retain the parent-band dispersions and appear only during pump--probe overlap. Rotating the pump's linear polarization at fixed probe polarization redistributes replica weight between momentum-space branches. With a common parameter set, a model accounting for the optical fields inside and outside the solid captures this redistribution through coherent Floquet--Volkov interference more closely than either pathway alone or their incoherent intensity sum. These findings extend Floquet dressing to the itinerant bulk bands of a multiband metal without relying on the observation of energy gaps opened by hybridization between parent bands and their photon-dressed replicas, providing a basis for exploring how coherent optical dressing couples to metallic transport and collective electronic phases.\\}

\noindent {\bf Introduction}

Periodic optical driving extends the spatial periodicity of a crystal into the time domain. Under a field with period $T=2\pi/\omega$, a Bloch electron can be described by a Floquet state with a time-periodic component and a quasienergy defined modulo $\hbar\omega$ \cite{JHShirley_PR1965_JHShirley,HSambe_PRA1973_HSambe}. Each equilibrium band is thereby accompanied by photon-dressed replicas. Hybridization between replicas of different photon order $n$, separated by integer multiples of $\hbar\omega$ in the extended-zone representation, can open dynamical gaps and reconstruct the electronic wavefunctions \cite{TOka_PRB2009_HAoki,TKitagawa_PRB2010_EDemler,NHLindner_NP2011_VGalitski,TKitagawa_PRB2011_EDemler,MSRudner_PRX2013_MLevin}. Because the driving field is optically tunable, Floquet engineering offers a reversible route to controlling band dispersions, symmetry and topology and to creating nonequilibrium quantum phases \cite{MBukov_AP2015_APolkovnikov,AEckardt_RMP2017_AEckardt,TOka_ARCM2019_SKitamura,MSRudner_NRP2020_NHLindner,DNBasov_NM2017_DHsieh,AdelaTorre_RMP2021_MASentef,MASentef_NC2015_TPDevereaux,HHubener_NC2017_ARubio,JHMentink_NC2015_MEckstein,JWMcIver_NP2020_ACavalleri}.

Time- and angle-resolved photoemission spectroscopy (tr-ARPES) directly probes these states in energy, momentum and time. Momentum-resolved Floquet--Bloch states were first observed in the Dirac surface state of Bi$_2$Se$_3$ \cite{YHWang_Science2013_NGedik}, where later polarization-dependent measurements revealed the interplay between Floquet--Bloch and Volkov photoemission pathways \cite{FMahmood_NP2016_NGedik}. Subsequent work addressed scattering \cite{SAeschlimann_NL2021_IGierz} and subcycle dynamics \cite{SIto_Nature2023_RHuber}, and extended coherent dressing to black phosphorus \cite{SZhou_Nature2023_SZhou,SZhou_PRL2023_SZhou,CBao_NC2024_SZhou,BFan_SA2025_PTang,CBao_PRB2025_SZhou}, excitonic and valley systems \cite{YKobayashi_NP2023_SGhimire,EJSie_Science2017_NGedik}, optical nonlinearities \cite{JYShan_Nature2021_DHsieh}, and 2H-WSe$_2$ \cite{SFragkos_NC2025_SBeaulieu}. Photon-spaced replicas alone, however, do not establish initial-state Floquet dressing. Hybridization gaps between replicas and parent bands provide a diagnostic, but require suitable band crossings and splittings exceeding the spectral linewidth \cite{YHWang_Science2013_NGedik,FWang_NM2026_SZhou}. Recent studies of monolayer graphene used polarization-dependent Floquet--Volkov interference to identify an initial-state contribution without resolving such gaps \cite{DChoi_NP2025_NGedik,MMerboldt_NP2025_SMathias}. These studies established an alternative criterion in a two-dimensional Dirac semimetal, while leaving open whether it remains effective across the overlapping bands of a bulk metal.

This extension is compelling because states near the Fermi surface govern transport and collective electronic phases. A related experiment on Cu(111) involved the comparatively isolated Shockley surface and image-potential states, which lie within a projected bulk band gap \cite{MReutzel_NC2020_HPetek}. In a gapless bulk continuum, by contrast, absorption, itinerant-electron scattering and decoherence broaden the spectrum and generate incoherent populations that overlap with the coherent response \cite{SASato_JPB2020_ARubio,KISeetharam_PRX2015_GRefael,TIadecola_PRB2015_CChamon}. Strong screening introduces a further difficulty: the measured sidebands can be dominated by laser-assisted photoemission (LAPE), in which the pump dresses the outgoing photoelectrons to produce Volkov sidebands \cite{LMiajaAvila_PRL2006_GSaathoff,MKeunecke_PRB2020_SMathias}. These final-state replicas share the photon-energy spacing, parent-band dispersion and temporal confinement expected of Floquet replicas \cite{STPark_PRA2014_STPark}. Identifying Floquet dressing in a metal therefore requires distinguishing an initial-state contribution within a photoemission response that may be largely determined by LAPE. The optical boundary conditions are central to this distinction. Fresnel reflection, transmission and screening make the fields acting inside the material and on the outgoing photoelectron different in amplitude, direction and phase \cite{MKeunecke_PRB2020_SMathias,MMerboldt_NP2025_SMathias,QCourtade_PRB2026_SBeaulieu}. A finite Volkov contribution can persist under $s$ polarization, defined perpendicular to the pump incidence plane, so a residual sideband at $s$ alone cannot establish initial-state dressing. The interference criterion must therefore be tested against the allowed final-state response, using complementary energy-, momentum-, time- and polarization-resolved evidence \cite{FBoschini_RMP2024_ADamascelli,KGadge_PRB2026_SRManmana}.

The layered metal 2H-NbSe$_2$ provides a stringent setting for exploring Floquet--Volkov physics. Its partially occupied Nb-derived manifold forms several overlapping Fermi-surface sheets, while the same itinerant states participate in charge-density-wave (CDW) order and superconductivity \cite{LFMattheiss_PRL1973_LFMattheiss,KRossnagel_PRB2001_JPollmann,TValla_PRL2004_HBerger,TKiss_NP2007_HTakagi,SVBorisenko_PRL2009_HBerger,DJRahn_PRB2012_KRossnagel,FWeber_PRB2018_VNStrocov,JAWilson_AP1975_SMahajan,DEMoncton_PRL1975_FJDiSalvo,MDJohannes_PRB2006_CAHowells,APasztor_NC2021_CRenner}. 
In the momentum region probed here, these metallic bands are separated from higher-lying unoccupied bands by a comparatively large energy interval, placing their expected first-order replicas within a sparse spectral window. This separation reduces spectral overlap and resonant mixing with higher-lying real states, as well as contamination by incoherent photoexcited populations. NbSe$_2$ thus combines a bulk multiband metallic state with a favourable window for detecting its weak replicas, suggesting a practical materials-selection strategy for Floquet spectroscopy in metals.

Here we use polarization-dependent Floquet-Volkov interference to identify an initial-state contribution in the itinerant bulk-derived bands of 2H-NbSe$_2$. Energy- and time-resolved measurements establish the photon-shifted, field-synchronous character of the replicas, while polarization scans reveal how their weight is redistributed across the complete two-dimensional momentum field of view. We compare this response with calculations that treat the internal and external optical fields separately to describe the initial- and final-state contributions. Applying one global parameter set to the integrated sideband yield and momentum distributions tests whether the initial-state Floquet contribution remains identifiable despite screening and overlapping metallic bands.\\

\noindent {\bf Results}

\begin{figure*}[tbp]
\begin{center}
\includegraphics[width=1\columnwidth,angle=0]{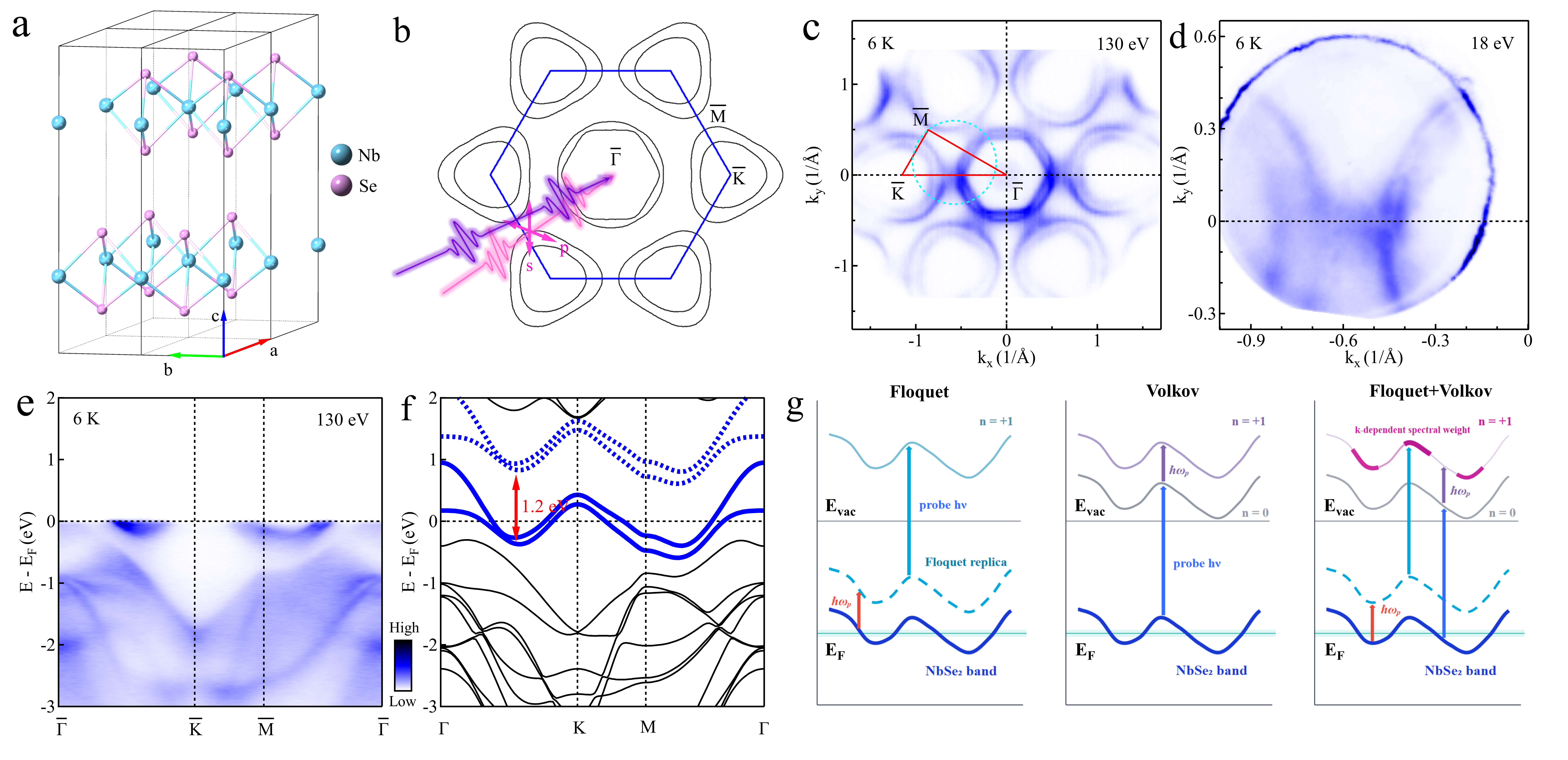}
\end{center}
\caption{\footnotesize\textbf{Equilibrium electronic structure and photon-dressed photoemission pathways in 2H-NbSe$_2$.} (a) Crystal structure of 2H-NbSe$_2$. (b) Calculated Fermi-surface contours in the (0001) surface Brillouin zone and a schematic of the pump--probe geometry. The polarization of the linearly polarized infrared (IR) pump can be rotated continuously between $s$ and $p$ polarization while the sample and photoelectron-detection geometry remain fixed. (c) Equilibrium Fermi-surface map measured with 130~eV photons. The cyan dashed circle marks the momentum region accessible with the 18~eV probe. (d) Fermi-surface map measured with the 18~eV probe. (e) Equilibrium band dispersion along the $\bar{\Gamma}$--$\bar{K}$--$\bar{M}$--$\bar{\Gamma}$ path marked in red in (c). (f) Calculated band structure with selected metallic bands highlighted in blue. The blue dashed lines denote their expected first-order replicas shifted upward by 1.2~eV into a comparatively sparse unoccupied-state window. (g) Schematics of initial-state Floquet--Bloch dressing, final-state Volkov dressing and their coherent combination. In the combined process, the two amplitudes reach the same detected $n=+1$ sideband and interfere to redistribute its momentum-dependent spectral weight.
}
\label{1}
\end{figure*}

To locate the expected replicas, we first consider the photon-shifted copies of the equilibrium dispersion. Neglecting drive-induced energy renormalization and hybridization, their energies are
\begin{equation}
E_n(\mathbf{k})=E_0(\mathbf{k})+n\hbar\omega_{\mathrm{pump}},
\label{eq:floquet_replica}
\end{equation}
where $n$ is an integer and $E_0(\mathbf{k})$ is the equilibrium band energy~\cite{JHShirley_PR1965_JHShirley}. The relevant experimental signature is therefore not simply photoexcited weight above the Fermi level, but a feature that retains the momentum dispersion of its parent band while being displaced by the pump photon energy. Identifying such a replica requires an equilibrium reference. We therefore examine 2H-NbSe$_2$, a layered multiband metal composed of weakly coupled Se--Nb--Se layers (Fig.~\ref{1}a).

The large-momentum-range ARPES map resolves pockets around the centre and corners of the (0001) surface Brillouin zone, whose hexagonal geometry is defined in Figs.~\ref{1}b,c. The 18~eV equilibrium map resolves the Fermi-surface contours within a smaller momentum region (Fig.~\ref{1}d), while the high-symmetry dispersion shows several bands crossing $E_{\mathrm F}$ (Fig.~\ref{1}e). These measurements establish both the multiband metallic character of 2H-NbSe$_2$ and the parent dispersions used below to identify the driven states. Although bulk 2H-NbSe$_2$ undergoes a CDW transition at $T_{\mathrm{CDW}}\approx33$~K~\cite{DEMoncton_PRL1975_FJDiSalvo,FWeber_PRL2011_DReznik}, no pronounced CDW-induced band folding is resolved within the sensitivity of our temperature-dependent measurements (Supplementary Information, Section~1). The directly observed metallic bands therefore provide the appropriate equilibrium reference.

The equilibrium band structure also provides a favourable window for resolving weak light-induced replicas. Shifting the selected metallic bands upward by $\hbar\omega_{\mathrm{pump}}=1.2$~eV places their expected $n=+1$ replicas in a comparatively sparse region of the unoccupied spectrum (Fig.~\ref{1}f), thereby reducing overlap with incoherently populated real states. This separation improves replica visibility, but it does not establish where the pump-photon exchange occurs. As summarized in Fig.~\ref{1}g, the pump may periodically dress the initial Bloch state inside the material, or the outgoing photoelectron may exchange a pump photon in the final state. Both pathways reach the same detected sideband and their amplitudes can therefore interfere, redistributing its spectral weight in momentum space. The equilibrium reference therefore identifies where replicas should appear, while the two pathways make their origin ambiguous. We first use the delay dependence to separate the field-synchronous response from longer-lived photoexcited populations.

\begin{figure*}[tbp]
\begin{center}
\includegraphics[width=1\columnwidth,angle=0]{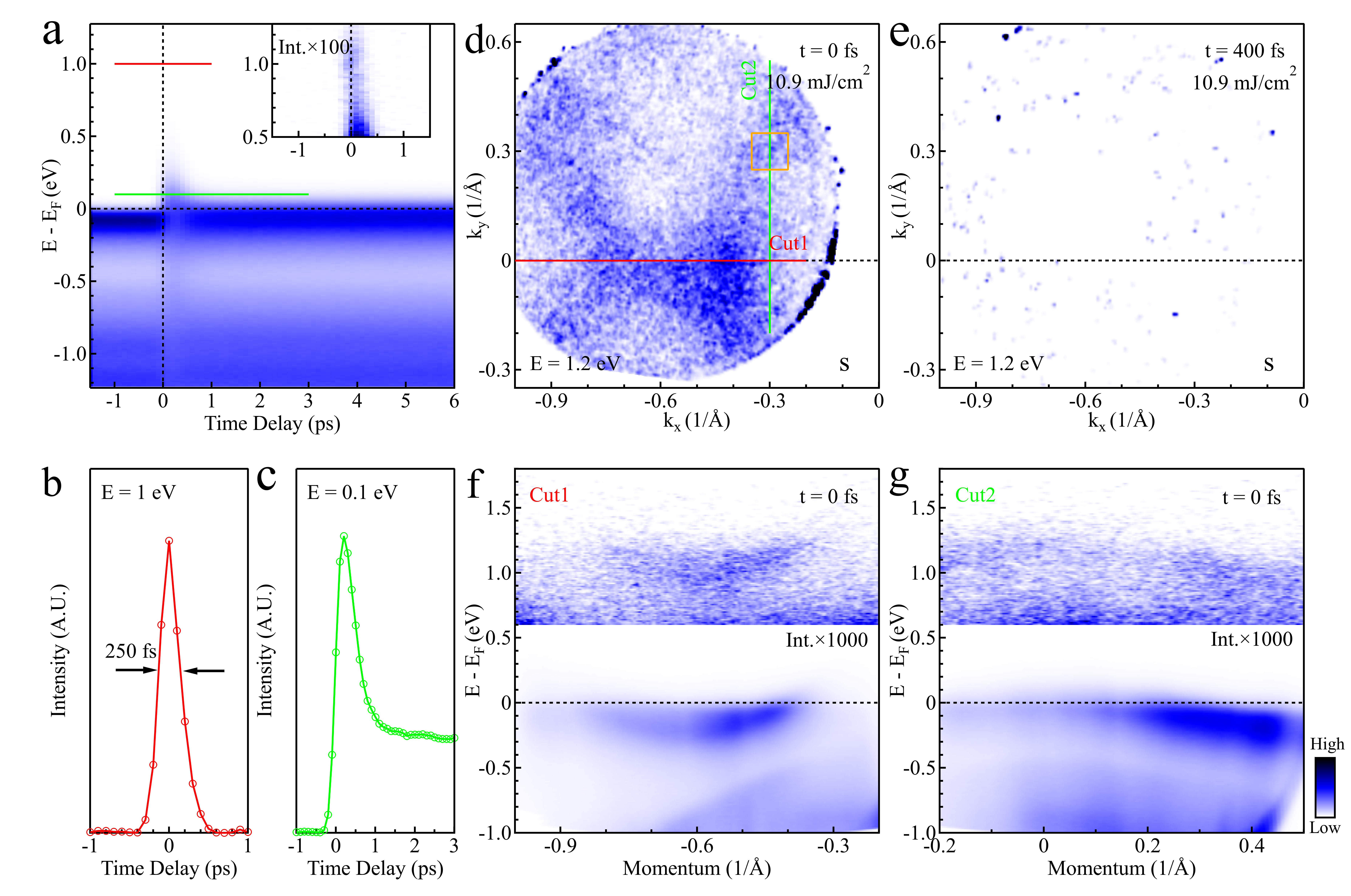}
\end{center}
\caption{\footnotesize\textbf{Temporal evolution of the photon-dressed replicas under 1.2~eV pumping.} (a) Photoemission intensity as a function of energy and pump--probe delay. The signal is integrated over the momentum region indicated by the orange box in (d). The red and green lines mark the energies used to extract the traces in (b) and (c), respectively. The inset magnifies the high-energy sideband region, with its intensity multiplied by 100. (b,c) Delay-dependent intensities at 1.0 and 0.1~eV, respectively. The approximately 250~fs width of the high-energy response is indicated in (b). (d,e) Constant-energy sideband maps at $E-E_{\mathrm F}=1.2$~eV measured at $t=0$ and 400~fs, respectively, with an $s$-polarized pump at an incident peak fluence of $10.9~\mathrm{mJ\,cm^{-2}}$. (f,g) Band dispersions measured at $t=0$~fs along Cut1 and Cut2 marked in (d), respectively. The intensity above 0.6~eV is multiplied by 1000 for visibility.
}
\label{2}
\end{figure*}

The delay dependence in Fig.~\ref{2} separates the two responses in time. The high-energy sideband is sharply confined to the pump--probe temporal overlap, whereas the photoexcited population close to $E_{\mathrm F}$ persists for several picoseconds (Fig.~\ref{2}a). The corresponding traces make this contrast explicit (Figs.~\ref{2}b,c): the sideband intensity forms a nearly symmetric peak centred at $t=0$, consistent with confinement to the temporal overlap of the pulses, while the low-energy population rises around $t=0$ and then relaxes slowly. The high-energy response therefore tracks the presence of the optical field rather than the lifetime of an incoherently populated state.

The momentum-resolved measurements show that this field-synchronous signal replicates the equilibrium electronic structure. At temporal overlap, its constant-energy contour reproduces the characteristic momentum dependence of the equilibrium Fermi surface (Fig.~\ref{2}d). Along both representative cuts, the additional high-energy features follow the dispersions of the occupied parent bands with an energy displacement equal to the pump photon energy, $\hbar\omega_{\mathrm{pump}}=1.2$~eV (Figs.~\ref{2}f,g).  Comparison with the static dispersion further shows that the occupied parent band is transiently broadened at $t=0$~fs, concomitant with the formation of the high-energy replicas (Supplementary Fig.~S3). Their resolved dispersions distinguish the replicas from a featureless photoexcited background. Once the pump and probe no longer overlap, the replicated contours disappear even though the low-energy population remains visible (Fig.~\ref{2}e). The absence of residual spectral weight in this window rules out a detectable long-lived excited bulk or surface population as the source of the sideband and is inconsistent with a static band or a purely thermal population. Fluence-dependent measurements show that the replica becomes resolvable as its intensity rises above the detection limit, with progressively stronger weight at higher pump fluence (Supplementary Information, Section~2).

The fixed photon-energy offset, reproduction of the parent-band dispersions and Fermi-surface contours, and confinement to the pump--probe overlap identify the replicas as a field-driven response rather than an incoherent excited-state population. These observations alone, however, cannot determine whether the dressing occurs in the initial Bloch states or in the outgoing photoelectrons, because Floquet--Bloch and Volkov processes share the same energy spacing and temporal confinement. We therefore use the polarization-dependent two-dimensional momentum texture to separate their contributions \cite{DChoi_NP2025_NGedik,MMerboldt_NP2025_SMathias}.

\begin{figure*}[tbp]
\begin{center}
\includegraphics[width=1\columnwidth,angle=0]{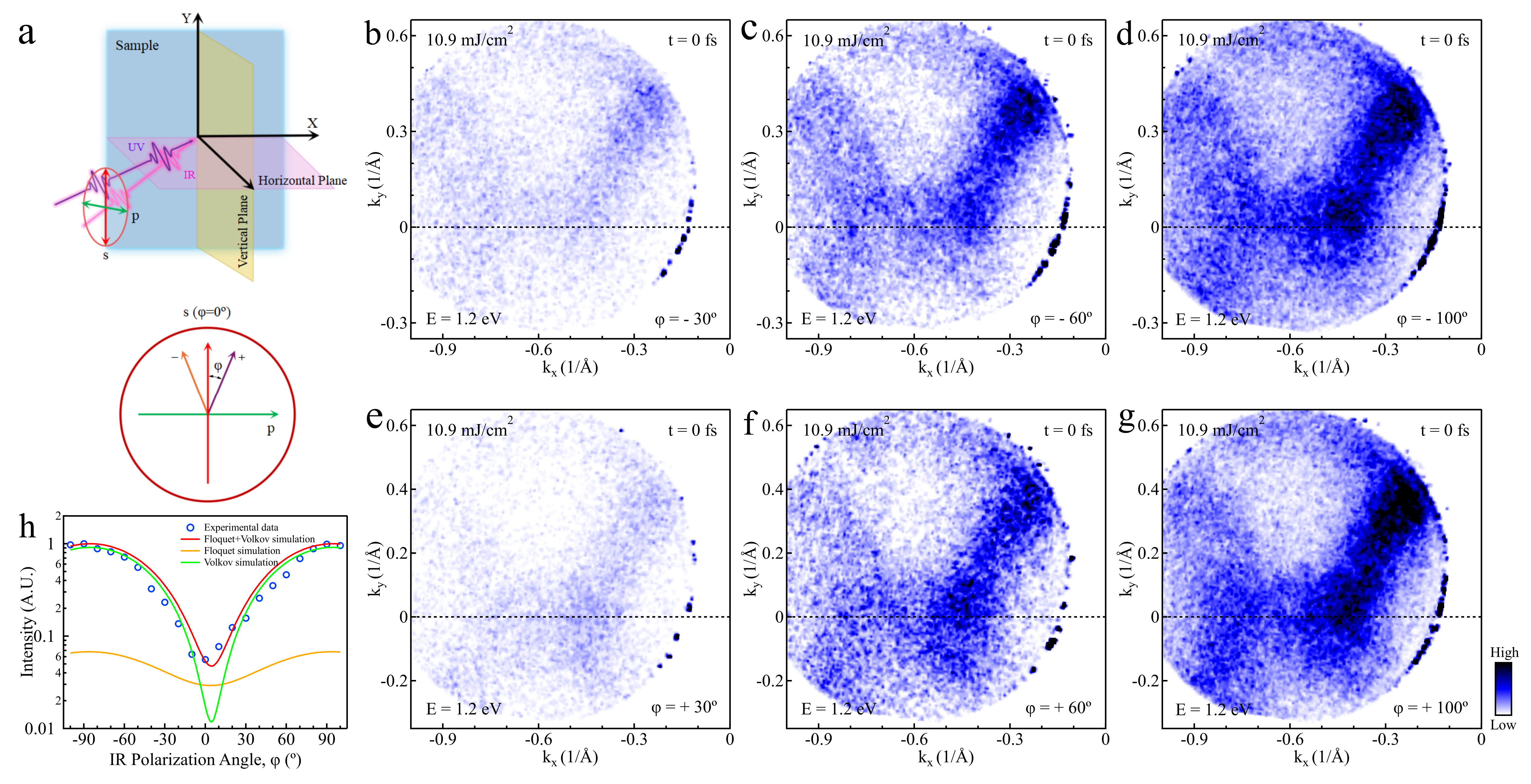}
\end{center}
\caption{\footnotesize\textbf{Pump-polarization dependence of the photon-dressed replicas.} (a) Non-collinear beam geometry and definition of the pump-polarization angle: $\varphi=0^\circ$ is $s$ polarization and $\varphi=\pm90^\circ$ is $p$ polarization, referenced to the pump incidence plane. The probe polarization is fixed. (b--d) Intensity-normalized constant-energy sideband maps at $E-E_{\mathrm F}=1.2$~eV and $t=0$~fs measured at $\varphi=-30^{\circ}$, $-60^{\circ}$ and $-100^{\circ}$, respectively. (e--g) Corresponding maps measured at $\varphi=+30^{\circ}$, $+60^{\circ}$ and $+100^{\circ}$, respectively. All maps were acquired at an incident pump peak fluence of $10.9~\mathrm{mJ\,cm^{-2}}$ and are displayed on the same normalized intensity scale.  At each angle, a single factor normalizes the intensity integrated over the full measured momentum region and $-1.0\leq E-E_{\mathrm F}\leq1.8$~eV to a common value (Supplementary Fig.~S4). (h) Normalized photoemission intensity integrated over the full measured momentum region and the energy window $0.6\leq E-E_{\mathrm F}\leq1.6$~eV as a function of pump-polarization angle. Blue open circles show the experimental data; the orange, green and red curves represent the Floquet-only, Volkov-only and coherently combined Floquet--Volkov calculations, respectively. All three calculated curves are divided by the maximum intensity of the coherent Floquet--Volkov response, setting its peak to unity while preserving their relative intensities. The vertical axis is logarithmic. The Volkov amplitude scale is calibrated in the coherent model to the measured $p$-to-$s$ contrast of 18.6 and then retained for all curves and angles.
}
\label{3}
\end{figure*}

The pump-polarization scan provides this additional constraint by rotating the driving field while keeping the sample, probe and detection geometry fixed (Fig.~\ref{3}a). It tests how pump polarization affects both the total sideband yield and the distribution of sideband intensity in momentum space.

Each pump-polarization dataset is scaled by a single factor so that its intensity, integrated over the full measured momentum region and $-1.0\leq E-E_{\mathrm F}\leq1.8$~eV, is the same at every angle (Methods; Supplementary Information, Section~4 and Fig.~S4). The resulting constant-energy maps show that rotating the pump polarization redistributes the sideband intensity across momentum space (Fig.~\ref{3}b--g). The intensity remains strongly anisotropic at all measured polarization angles, and reversing the sign of $\varphi$ changes the relative intensities in different momentum regions rather than uniformly rescaling the entire contour.

\begin{figure*}[tbp]
\begin{center}
\includegraphics[width=1\columnwidth,angle=0]{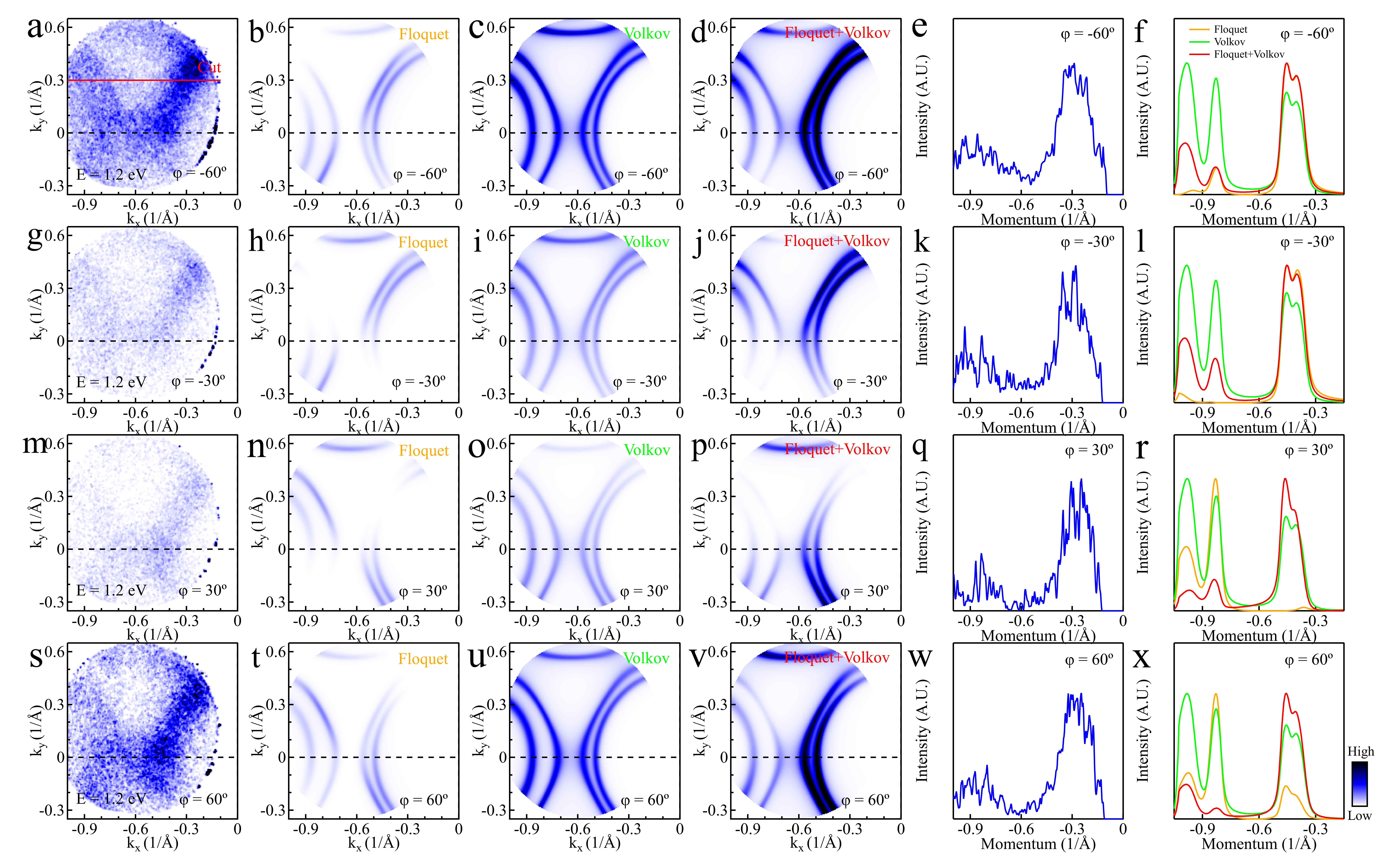}
\end{center}
\caption{\footnotesize\textbf{Momentum-resolved comparison of experiment and Floquet--Volkov calculations.} (a--f) Results for $\varphi=-60^{\circ}$. (a) Experimental constant-energy sideband map at $E-E_{\mathrm F}=1.2$~eV and $t=0$~fs. The red line marks the momentum cut used to extract the momentum distribution curves (MDCs). (b--d) Corresponding Floquet-only, Volkov-only and coherently combined Floquet--Volkov calculations, respectively. (e) Experimental MDC along the cut in (a), shown in blue. (f) Corresponding calculated MDCs along the same cut for the Floquet-only (orange), Volkov-only (green) and coherent Floquet--Volkov (red) responses, all normalized by the maximum intensity across the three curves. (g--l), (m--r) and (s--x) Same as (a--f), but for $\varphi=-30^{\circ}$, $+30^{\circ}$ and $+60^{\circ}$, respectively, using the same momentum cut. The experimental maps in (a,g,m,s) are replotted from Fig.~\ref{3}c,b,e,f, respectively, to facilitate comparison with the calculations. The photoemission model and MDC extraction are described in Supplementary Information, Section~9.
}
\label{4}
\end{figure*}

Figure~\ref{3}h summarizes the momentum-integrated sideband yield over the full measured momentum region and the energy window $0.6\leq E-E_{\mathrm F}\leq1.6$~eV. The measured intensity has a finite minimum near $s$ polarization and increases towards $p$ polarization. Residual intensity near $\varphi=0^\circ$ is not by itself evidence for initial-state Floquet dressing: an incoherent photoexcited background, imperfect linear polarization or a finite Volkov response can also fill the minimum. The relative Volkov scale is calibrated to the measured $p$-to-$s$ intensity contrast (Supplementary Information, Section~9.5), so this contrast is a calibration input rather than independent evidence for a dressing pathway. With one common parameter set, the coherent Floquet--Volkov calculation follows the intermediate-angle evolution most closely. Because background and normalization affect the depth of the minimum, the integrated yield alone does not identify the initial-state contribution. We therefore turn to the branch-resolved momentum distributions in Fig.~\ref{4}.


With the relative scale constrained by the integrated yield (Fig.~\ref{3}h), the momentum distributions provide a more stringent test of the two photoemission pathways. We compare the experimental maps at intermediate polarization angles (Fig.~\ref{4}a,g,m,s) with the Floquet-only, Volkov-only and coherent Floquet--Volkov responses, retaining the same model parameters throughout. The comparison now tests how the sideband weight is distributed between distinct branches, information that is lost in the momentum integral.

The two pathways couple differently to momentum and pump polarization because they involve electrons inside the solid and photoelectrons in the near-surface vacuum. Our model accounts for the effective fields in these regions through Fresnel calculations using the complex dielectric tensor of NbSe$_2$ (Supplementary Information, Section~9). Their coherent superposition introduces an interference term that can enhance or suppress the intensity in different momentum positions. The Floquet-only and Volkov-only maps predict distinct branch intensities (Fig.~\ref{4}b,c,h,i,n,o,t,u), while their coherent combination more closely captures the dominant branches and their polarization-dependent redistribution (Fig.~\ref{4}d,j,p,v).

The momentum distribution curves (MDCs) along the cut marked in Fig.~\ref{4}a make this distinction explicit. The experimental MDCs remain dominated by the right-hand feature at all four polarization angles (Fig.~\ref{4}e,k,q,w), whereas the Volkov-only calculation assigns excessive weight to the left-hand features (Fig.~\ref{4}f,l,r,x). This discrepancy in relative spectral weight cannot be removed by an overall intensity rescaling. The Floquet-only response also fails, favouring the left-hand features at positive polarization angles. Coherently combining the two amplitudes suppresses the left-hand weight and best reproduces the measured branch imbalance. Simply adding the Floquet-only and Volkov-only intensities retains excessive left-hand weight and fails to reproduce this imbalance (Supplementary Information, Section~6 and Fig.~S6). Within the same effective-field model and parameter set, these comparisons require a finite initial-state Floquet--Bloch contribution and its interference with the Volkov amplitude to account for the observed momentum distribution. The calculated polarization-dependent momentum texture also persists when the relative Volkov scale is varied by $\pm20\%$ with all other parameters fixed (Supplementary Information, Section~9.7 and Fig.~S9).

At $s$, by contrast, the comparable left- and right-hand spectral weights are reproduced by all three calculations, so this profile alone does not distinguish the pathways. The model nevertheless predicts a predominantly Floquet response with a weaker, finite Volkov contribution (Supplementary Information, Section~7 and Fig.~S7). Small rotations away from $s$ polarization redistribute weight in both the constant-energy maps and the dispersing replica, consistent with the coherent response (Supplementary Information, Section~8 and Fig.~S8). A complementary polarization-reversal asymmetry analysis gives the same ordering of agreement between the three calculated responses (Supplementary Information, Section~5 and Fig.~S5). Together, these polarization-dependent measurements support the coherent Floquet--Volkov interpretation.

Applying this interferometric criterion to 2H-NbSe$_2$ extends its use from the Dirac states of monolayer graphene to overlapping bulk-derived metallic bands \cite{DChoi_NP2025_NGedik,MMerboldt_NP2025_SMathias}, where screening, finite linewidths and incoherent carriers obscure the weak replicas. Their separation from higher-lying real states provides a spectral window for detection (Fig.~\ref{1}f), while the delay and polarization dependences distinguish the field-synchronous response and constrain its initial- and final-state contributions (Figs.~\ref{2}--\ref{4}). Resolving this response is aided by the time-of-flight (TOF) analyser, which simultaneously records the complete two-dimensional momentum distribution over a broad energy window at the single-electron-event level \cite{QDGuo_SD2022_OTjernberg,JMaklar_RSI2020_LRettig}. This parallel acquisition preserves the connection between the weak sideband and its contour-dependent polarization response without assembling the momentum pattern from sequential cuts.


In summary, polarization-resolved photoemission identifies an initial-state Floquet contribution in the bulk metallic bands of 2H-NbSe$_2$ through its interference with the final-state Volkov response. The momentum-dependent redistribution supplies information that is absent from the total sideband yield and remains accessible when a light-induced gap is unresolved. This extends Floquet--Volkov interferometry to a screened multiband metal and offers a basis for examining how coherent electronic dressing couples to transport, CDW order and superconductivity.\\


\noindent {\bf Methods}\\
\noindent{\bf Sample growth} Single crystals of 2H-NbSe$_2$ were grown by chemical vapour transport (CVT). Stoichiometric amounts of niobium powder and selenium pieces, together with iodine at $1~\mathrm{mg\,ml^{-1}}$ as the transport agent, were sealed in an evacuated quartz ampoule. The ampoule was heated gradually and held under a temperature gradient from $850\,^{\circ}\mathrm{C}$ to $670\,^{\circ}\mathrm{C}$ for two weeks, before being cooled naturally to room temperature. Millimetre-sized crystals (approximately 3~mm) with regular shapes and shiny surfaces were obtained.

\noindent{\bf ARPES measurements} The tr-ARPES measurements were performed using a high-harmonic-generation (HHG)-based system at BALTAZAR, KTH. The system combines 18~eV probe photons with a TOF electron energy analyser, enabling simultaneous acquisition over a two-dimensional momentum field of view in $(k_x,k_y)$ space \cite{QDGuo_SD2022_OTjernberg}. The samples were measured at approximately 6~K under ultrahigh-vacuum conditions with a base pressure below $1.0\times10^{-10}$~mbar. The pump pulses had a wavelength of 1030~nm, corresponding to a photon energy of $\hbar\omega_{\mathrm{pump}}=1.2$~eV, a repetition rate of 250~kHz and a pulse duration of approximately 250~fs. The pump spot on the sample had full widths at half maximum (FWHM) of approximately $130\times130~\mu\mathrm{m}^2$, and the average pump power was varied from 50~mW to 520~mW. For a two-dimensional Gaussian pump profile, we report the incident on-axis peak fluence as $F_{\mathrm{peak}}=4\ln 2\,P/(\pi f_{\mathrm{rep}}d_xd_y)$, where $P$ is the average pump power, $f_{\mathrm{rep}}$ is the repetition rate and $d_x,d_y$ are the spot FWHM diameters. At the maximum power, this gives $F_{\mathrm{peak}}=10.9~\mathrm{mJ\,cm^{-2}}$. All fluences in the main text and Supplementary Information use this peak-fluence convention. The 18~eV probe pulses operated at the same repetition rate of 250~kHz and had a spot size of approximately $90\times80~\mu\mathrm{m}^2$ (FWHM). Complementary equilibrium ARPES measurements were carried out at the I05 beamline of Diamond Light Source and at the Synergetic Extreme Condition User Facility (SECUF).

\noindent{\bf Pump and probe geometry} The pump incidence plane is horizontal in the laboratory frame. We define $s$ polarization perpendicular to this plane and $p$ polarization within it; these correspond to the laboratory labels linear-vertical (LV) and linear-horizontal (LH), respectively. The pump-polarization angle is $\varphi=0^\circ$ for $s$ and $\varphi=\pm90^\circ$ for $p$. The probe polarization remains fixed during the pump-polarization scan. The beams are non-collinear: the probe lies approximately $15^\circ$ below the horizontal plane, and its horizontal projection differs from the pump direction by approximately $18.5^\circ$, giving a beam separation of about $23^\circ$. This inter-beam separation is distinct from the pump incidence angle of $22^\circ$ relative to the surface normal used in the Fresnel calculation (Supplementary Information, Section~9.2).

\noindent{\bf Intensity normalization}
For each pump-polarization angle, the photoemission spectrum at temporal overlap is integrated over the full detected momentum region to obtain an energy-distribution curve (EDC). The EDC area over $-1.0\leq E-E_{\mathrm F}\leq1.8$~eV determines a single multiplicative factor that brings this area to the same value for all angles. This factor is applied uniformly to the corresponding energy--momentum dataset. The normalized data yield the constant-energy maps in Fig.~\ref{3}b--g and, after integration over the full measured momentum region and $0.6\leq E-E_{\mathrm F}\leq1.6$~eV, the sideband yield in Fig.~\ref{3}h. Details and representative EDCs are given in Supplementary Information, Section~4 and Fig.~S4.

\noindent{\bf Density functional theory calculations}
The electronic structure of 2H-NbSe$_2$ was calculated within density functional theory (DFT) using the Vienna \textit{Ab initio} Simulation Package (VASP) \cite{GKresse_PRB1993_JHafner,GKresse_PRB1996_JFurthmuller,GKresse_CMS1996_JFurthmuller}. The projector-augmented-wave (PAW) method was used with a plane-wave energy cutoff of 420~eV \cite{PEBlochl_PRB1994_PEBlochl,GKresse_PRB1999_DJoubert}. Exchange--correlation effects were treated within the generalized-gradient approximation using the Perdew--Burke--Ernzerhof functional \cite{JPPerdew_PRL1996_MErnzerhof}. The Brillouin zone was sampled with a $16\times16\times4$ $\Gamma$-centred k-point mesh. The crystal structure was relaxed until the Hellmann--Feynman force on each atom was below $10^{-2}~\mathrm{eV}\,\text{\AA}^{-1}$, with an energy-convergence criterion of $10^{-7}$~eV. Spin--orbit coupling was included in the electronic-structure calculations.

\noindent{\bf Photoemission calculations}
The polarization-dependent sideband is modelled using two Nb-$4d$-derived bands and the coherent sum of the initial-state Floquet--Bloch and final-state Volkov amplitudes. Separate effective fields inside the material and in the near-surface vacuum account for Fresnel transmission, reflection and dielectric screening. Floquet-only, Volkov-only and coherent distributions are evaluated with one global parameter set for all pump-polarization angles, and MDCs are extracted along the cut used for the experimental comparison. An incoherent control is obtained by adding the Floquet-only and Volkov-only intensities with the same parameters. The amplitude scale $C_{\mathrm V}$ is selected by matching the calculated coherent $p$-to-$s$ integrated intensity ratio to the measured ratio of 18.6, with the other parameters fixed. It is not an independently fitted intensity prefactor for the Volkov-only curve. Once calibrated, the same value is retained in the coherent, isolated-channel and incoherent-sum comparisons. The formalism, calibration definition, optical boundary conditions, model parameters and sensitivity analysis are described in Supplementary Information, Section~9. The Python implementation of the Floquet--Volkov model was developed with assistance from GPT-6 Astra (OpenAI). The authors reviewed and tested the code and take full responsibility for the calculations and their interpretation.

\noindent {\bf Data availability}\\
The data supporting the findings of this study are available within the Article and its Supplementary Information.

\vspace{3mm}

\noindent {\bf Acknowledgements}\\
Complementary measurements were performed at beamline I05 of Diamond Light Source under proposals SI39652 and SI41048a and at the D5 station of Synergetic Extreme Condition User Facility (SECUF), Institute of Physics, Chinese Academy of Sciences. We acknowledge both facilities for access and experimental support.

\noindent {\bf Funding}\\
C.L., M.H.B., W.Y.C. and O.T. acknowledge support from the Swedish Research Council (grants 2019-00701 and 2019-03486) and the Knut and Alice Wallenberg Foundation (grant 2018.0104). Q.R.C. and A.D. acknowledge support from the Swedish Research Council (grant 2024-04986), the Knut and Alice Wallenberg Foundation (grant 2022.0108), and the Göran Gustafsson Foundation. The computations were enabled by the National Academic Infrastructure for Supercomputing in Sweden (NAISS), partially funded by the Swedish Research Council through grant 2022-06725. M.D. acknowledges financial support from the Göran Gustafsson Foundation, Swedish Research Council under Grant No. 2022.03813 and the Carl Trygger Foundation. Z.L.L. acknowledges support from the Youth Innovation Promotion Association of the Chinese Academy of Sciences (grant 2021008).

\vspace{3mm}

\noindent {\bf Author Contributions}\\
C.L. proposed and conceived the project. C.L. carried out the ARPES experiments with assistance from M.D., M.H.B. and W.Y.C.. Q.R.C. and A.D. performed the band-structure calculations and theoretical model analysis. Z.L.L. grew the 2H-NbSe$_2$ single crystals. C.L. developed the data-analysis software, analysed the data and wrote the manuscript with assistance from Q.R.C.. C.L., Q.R.C., M.D., M.H.B., A.D., W.T.Z. and O.T. participated in scientific discussions. All authors commented on the manuscript.

\noindent {\bf Competing Interests}\\
The authors declare no competing interests.

\newpage

\end{document}


\centerline{\large{Supplementary Information for}}

\title{Floquet dressing of bulk metallic bands in 2H-NbSe$_2$}

\author{Cong Li$^{1,2,\sharp,*}$, Qirui Cui$^{2,\sharp,*}$, Maciej Dendzik$^{2}$, Magnus H. Berntsen$^{2}$, Wanyu Chen$^{2}$, Zhilin Li$^{1}$, Anna Delin$^{2}$, Wentao Zhang$^{1,*}$, Oscar Tjernberg$^{2,*}$
}

\affiliation{
\\$^{1}$Beijing National Laboratory for Condensed Matter Physics, Institute of Physics, Chinese Academy of Sciences, Beijing 100190, China
\\$^{2}$Department of Applied Physics, KTH Royal Institute of Technology, Stockholm 11419, Sweden
\\$^{\sharp}$These authors contributed equally to this work.
\\$^{*}$Corresponding authors: licong@iphy.ac.cn, qiruic@kth.se, wentaozhang@iphy.ac.cn, oscar@kth.se
}

\pacs{}

\maketitle

{\bf 1. Temperature dependence of the equilibrium electronic structure}

\begin{figure*}[tbp]
\begin{center}
\includegraphics[width=0.85\textwidth]{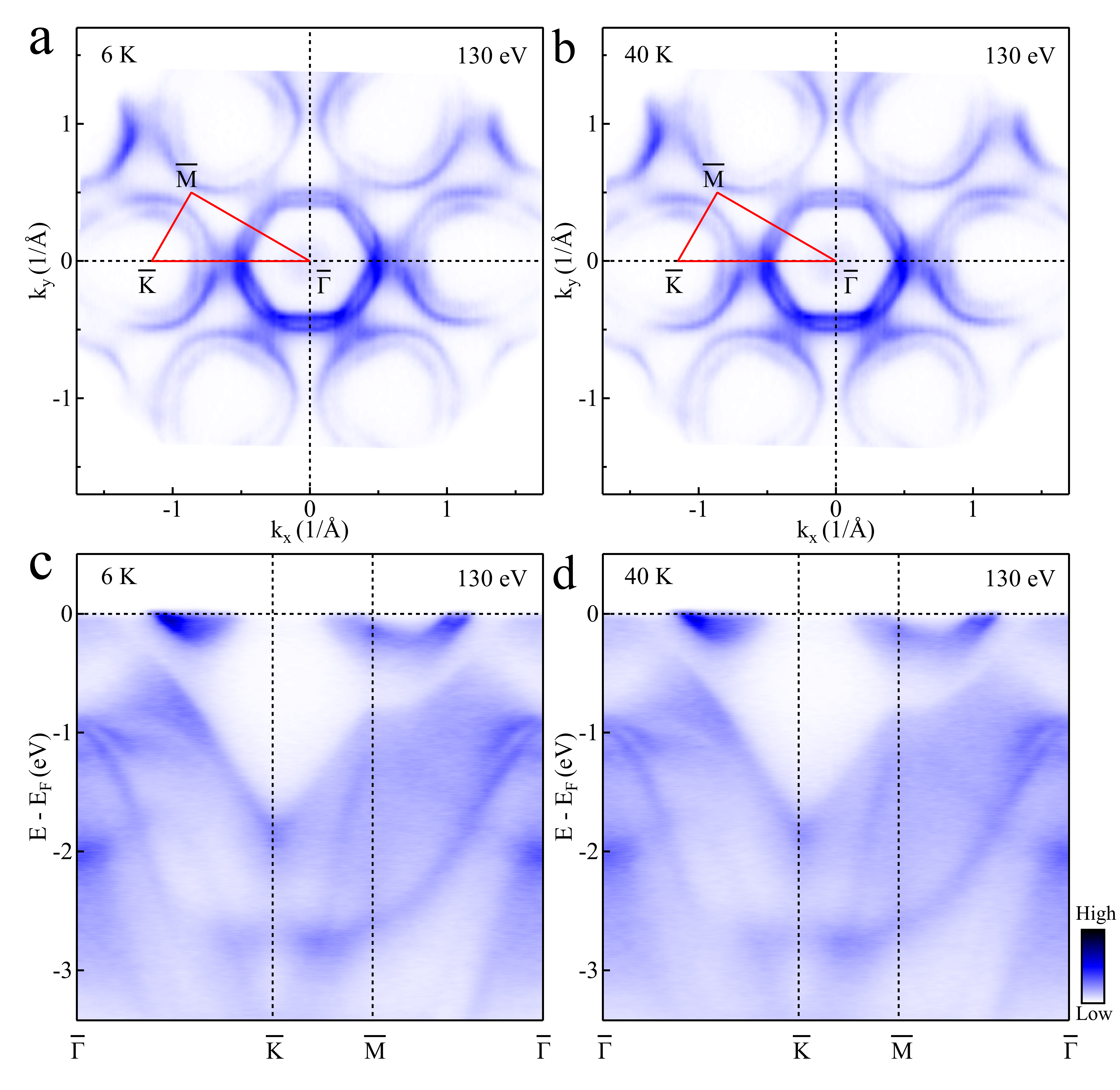}
\end{center}
\caption{\textbf{Temperature dependence of the equilibrium electronic structure of 2H-NbSe$_2$.}
(a,b) Fermi-surface maps measured with 130~eV photons at 6 and 40~K, respectively, straddling the reported $T_{\mathrm{CDW}}\approx33$~K~\cite{DEMoncton_PRL1975_FJDiSalvo,FWeber_PRL2011_DReznik}. The red lines indicate the $\bar{\Gamma}$-$\bar{K}$-$\bar{M}$-$\bar{\Gamma}$ momentum trajectory. (c,d) Corresponding angle-resolved photoemission spectroscopy (ARPES) dispersions measured along the trajectory indicated in (a) and (b).}
\label{fig:S1}
\end{figure*}

Temperature-dependent measurements test whether charge-density-wave (CDW) order substantially alters the equilibrium bands used to identify the photon-dressed replicas. Bulk 2H-NbSe$_2$ has a reported CDW transition temperature $T_{\mathrm{CDW}}\approx33$~K~\cite{DEMoncton_PRL1975_FJDiSalvo,FWeber_PRL2011_DReznik}. The Fermi-surface maps at 6 and 40~K, below and above this literature transition temperature, show similar resolved contours (Supplementary Fig.~\ref{fig:S1}a,b).

The corresponding dispersions show no pronounced CDW-induced band folding within the present sensitivity and resolution (Supplementary Fig.~\ref{fig:S1}c,d). We therefore use the experimentally resolved metallic bands as the reference for the pump-induced replicas, without assuming that small CDW gaps or weak folded features are absent.\\

{\bf 2. Pump-fluence dependence of the photon-dressed replicas}

\begin{figure*}[tbp]
\begin{center}
\includegraphics[width=\textwidth]{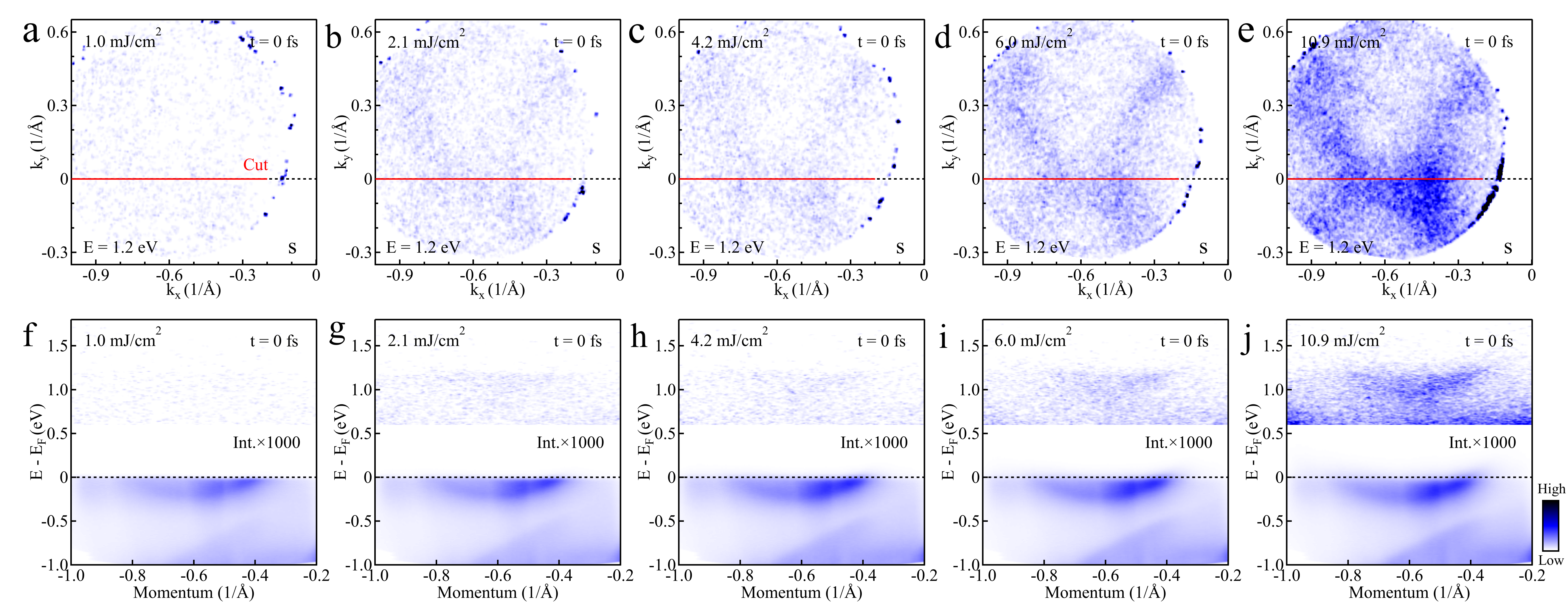}
\end{center}
\caption{\textbf{Pump-fluence dependence of the photon-dressed replicas.}
(a--e) Constant-energy contours at $E-E_{\mathrm{F}}=1.2$~eV and $t=0$~fs measured with $s$ pump polarization at incident peak fluences of 1.0, 2.1, 4.2, 6.0 and 10.9~$\mathrm{mJ\,cm^{-2}}$, respectively. The experimental geometry remains fixed throughout the fluence scan. The red lines indicate the momentum trajectory used for the band dispersions. (f--j) Corresponding band dispersions measured along the trajectories indicated in (a--e). The intensity above 0.6~eV is multiplied by 1000 for visibility.}
\label{fig:S2}
\end{figure*}

The fluence dependence provides an additional check on the pump-induced origin of the high-energy sideband. Under $s$ pumping, its constant-energy contours become progressively clearer with increasing fluence (Supplementary Fig.~\ref{fig:S2}a--e), whereas no discernible signal is observed at the lowest peak fluence of $1.0~\mathrm{mJ\,cm^{-2}}$. All values are incident peak fluences for a Gaussian pump profile, as defined in the main-text Methods.

The corresponding energy--momentum spectra along the trajectory indicated in the constant-energy maps are presented in Supplementary Fig.~\ref{fig:S2}f--j. The dispersive sideband between approximately 0.6 and 1.2~eV strengthens systematically with increasing pump fluence, while its characteristic dispersion and energy separation from the parent band remain essentially unchanged. This systematic enhancement supports the pump-induced origin of the sideband. The absence of a discernible signal at low fluence sets a practical detection limit under the present acquisition conditions, rather than establishing a sharp threshold for light dressing. The high-fluence measurements provide sufficient sideband spectral weight to resolve its complete two-dimensional momentum distribution and polarization dependence.\\

{\bf 3. Pump-induced spectral broadening of the occupied band}

\begin{figure*}[tbp]
\begin{center}
\includegraphics[width=\textwidth]{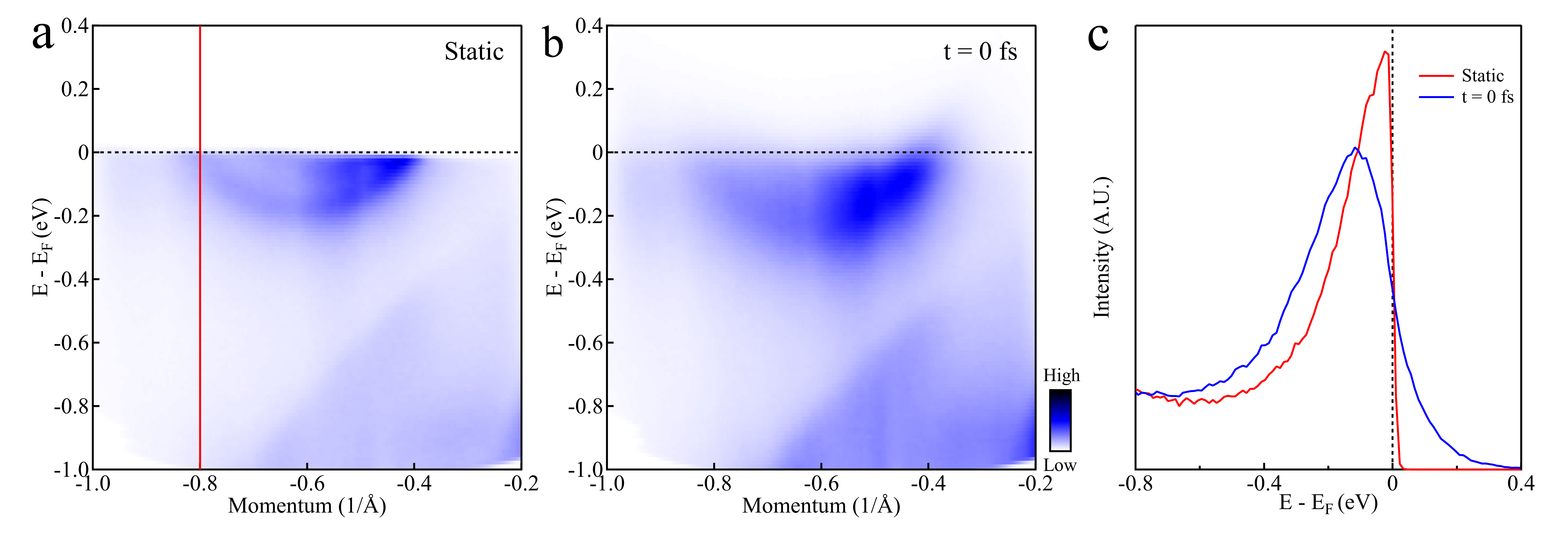}
\end{center}
\caption{\textbf{Pump-induced spectral broadening of the occupied band at temporal overlap.}
(a) Static band dispersion measured along the same momentum trajectory as Fig.~2f of the main text. The red line marks the momentum at which the energy-distribution curves (EDCs) are extracted. (b) Corresponding band dispersion measured at $t=0$~fs under pump excitation. (c) Comparison of the EDCs extracted from (a) and (b) at the indicated momentum. The static and $t=0$ spectra are shown in red and blue, respectively.}
\label{fig:S3}
\end{figure*}

At temporal overlap, the occupied parent band remains recognizable but develops a broader spectral profile along the same momentum trajectory used to resolve the high-energy replica in main-text Fig.~2f (Supplementary Fig.~\ref{fig:S3}a,b). The energy-distribution curves (EDCs), extracted at the same fixed momentum from the unpumped and pumped spectra, show an apparent peak shift and enhanced spectral weight across $E_{\mathrm F}$ under pumping (Supplementary Fig.~\ref{fig:S3}c). The replicas therefore coexist with a substantial modification of the parent-band spectrum. The EDC width alone does not determine the intrinsic quasiparticle lifetime, because the measured lineshape also depends on the occupation and experimental broadening.\\

{\bf 4. Normalization of the pump-polarization series}

\begin{figure*}[tbp]
\begin{center}
\includegraphics[width=\textwidth]{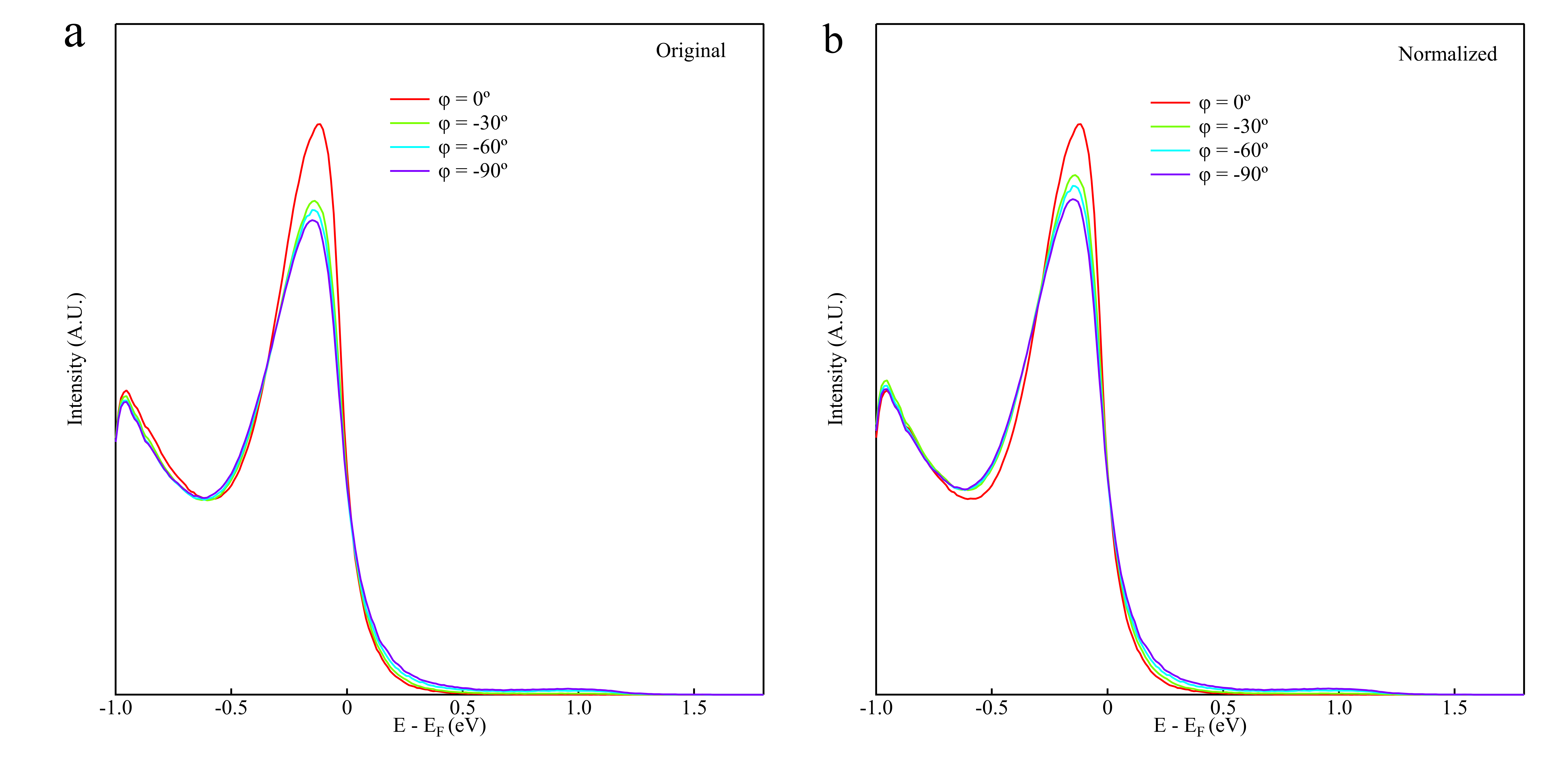}
\end{center}
\caption{\textbf{Normalization of the pump-polarization series.}
(a) Original energy-distribution curves (EDCs) obtained by integrating the photoemission intensity at temporal overlap over the full detected momentum region for pump-polarization angles $\varphi=-90^{\circ}$ (purple), $-60^{\circ}$ (cyan), $-30^{\circ}$ (green) and $0^{\circ}$ (red). (b) Corresponding EDCs after multiplication by an angle-dependent factor chosen to give the same integrated intensity over $-1.0\leq E-E_{\mathrm F}\leq1.8$~eV. The resulting factor is applied uniformly to the entire energy--momentum dataset at each angle to obtain the normalized maps and integrated sideband yields in main-text Fig.~3. The procedure is described in Section~4 and Eqs.~\eqref{eq:S_integrated_edc}--\eqref{eq:S_normalized_yield}.}
\label{fig:S4}
\end{figure*}

The pump-polarization series in Fig.~3 of the main text is normalized using the photoemission intensity integrated over the full measured momentum region and a fixed energy interval. At each pump-polarization angle $\varphi$, the spectrum at temporal overlap is first integrated over all detected momenta to obtain an energy-distribution curve (EDC),
\begin{equation}
D_{\varphi}(E)=\int_{\mathcal{K}} I_{\varphi}^{\mathrm{raw}}(\mathbf{k}_{\parallel},E)\,\mathrm{d}^{2}k_{\parallel},
\label{eq:S_integrated_edc}
\end{equation}
where $\mathcal{K}$ denotes the measured momentum region and $E$ is measured relative to $E_{\mathrm F}$. The area under this EDC is then evaluated over the same energy window at every angle,
\begin{equation}
A_{\varphi}=\int_{-1.0\,\mathrm{eV}}^{1.8\,\mathrm{eV}}D_{\varphi}(E)\,\mathrm{d}E.
\label{eq:S_normalization_area}
\end{equation}
The angle-dependent normalization factor is $N_{\varphi}=A_{\mathrm{ref}}/A_{\varphi}$, where $A_{\mathrm{ref}}$ is a common target area that sets the overall intensity scale. This single factor is applied to the entire energy--momentum dataset at the corresponding angle:
\begin{equation}
\begin{aligned}
I_{\varphi}^{\mathrm{norm}}(\mathbf{k}_{\parallel},E)&=N_{\varphi}I_{\varphi}^{\mathrm{raw}}(\mathbf{k}_{\parallel},E),\\
D_{\varphi}^{\mathrm{norm}}(E)&=N_{\varphi}D_{\varphi}(E).
\end{aligned}
\label{eq:S_normalized_intensity}
\end{equation}
Consequently, all normalized EDCs have the same integrated intensity between $-1.0$ and $1.8$~eV. Supplementary Fig.~\ref{fig:S4} illustrates the procedure for $\varphi=-90^{\circ}$, $-60^{\circ}$, $-30^{\circ}$ and $0^{\circ}$, showing the original and normalized EDCs in panels (a) and (b), respectively. Equal integrated areas do not require equal peak heights or identical lineshapes.

The constant-energy maps in main-text Fig.~3b--g are taken from the normalized datasets. The sideband yield in Fig.~3h is subsequently obtained by integrating these datasets over $\mathcal{K}$ and the narrower energy interval $0.6\leq E\leq1.6$~eV:
\begin{equation}
Y_{\varphi}=\int_{0.6\,\mathrm{eV}}^{1.6\,\mathrm{eV}}\int_{\mathcal{K}}
I_{\varphi}^{\mathrm{norm}}(\mathbf{k}_{\parallel},E)\,\mathrm{d}^{2}k_{\parallel}\,\mathrm{d}E.
\label{eq:S_normalized_yield}
\end{equation}
The normalization therefore preserves the relative momentum-dependent intensity within each map, while expressing the sideband yield relative to the spectral weight in the broader normalization window. It does not force the sideband intensities or EDC peak heights to be equal at different pump-polarization angles.\\

{\bf 5. Polarization-reversal asymmetry and comparison of photoemission pathways}

The momentum maps and momentum distribution curves (MDCs) in Fig.~4 of the main text compare the measured sideband weight with the Floquet-only, Volkov-only and coherent responses. Here we complement these comparisons with the normalized asymmetry between opposite pump-polarization angles,
\begin{equation}
A_{\varphi}(\mathbf{k}_{\parallel})=
\frac{I_{-\varphi}(\mathbf{k}_{\parallel})-I_{+\varphi}(\mathbf{k}_{\parallel})}
{I_{-\varphi}(\mathbf{k}_{\parallel})+I_{+\varphi}(\mathbf{k}_{\parallel})},
\label{eq:S_asymmetry}
\end{equation}
where $I_{\pm\varphi}$ denotes the normalized sideband intensity obtained using the procedure in Section~4 at $E-E_{\mathrm F}=1.2$~eV and temporal overlap. A common multiplicative intensity factor cancels in this ratio. Different overall yields at opposite angles can still contribute to the asymmetry, but a uniform rescaling alone would give a momentum-independent value. Its spatial variation therefore tests the redistribution of weight within the replica contours.

\begin{figure*}[tbp]
\begin{center}
\includegraphics[width=\textwidth]{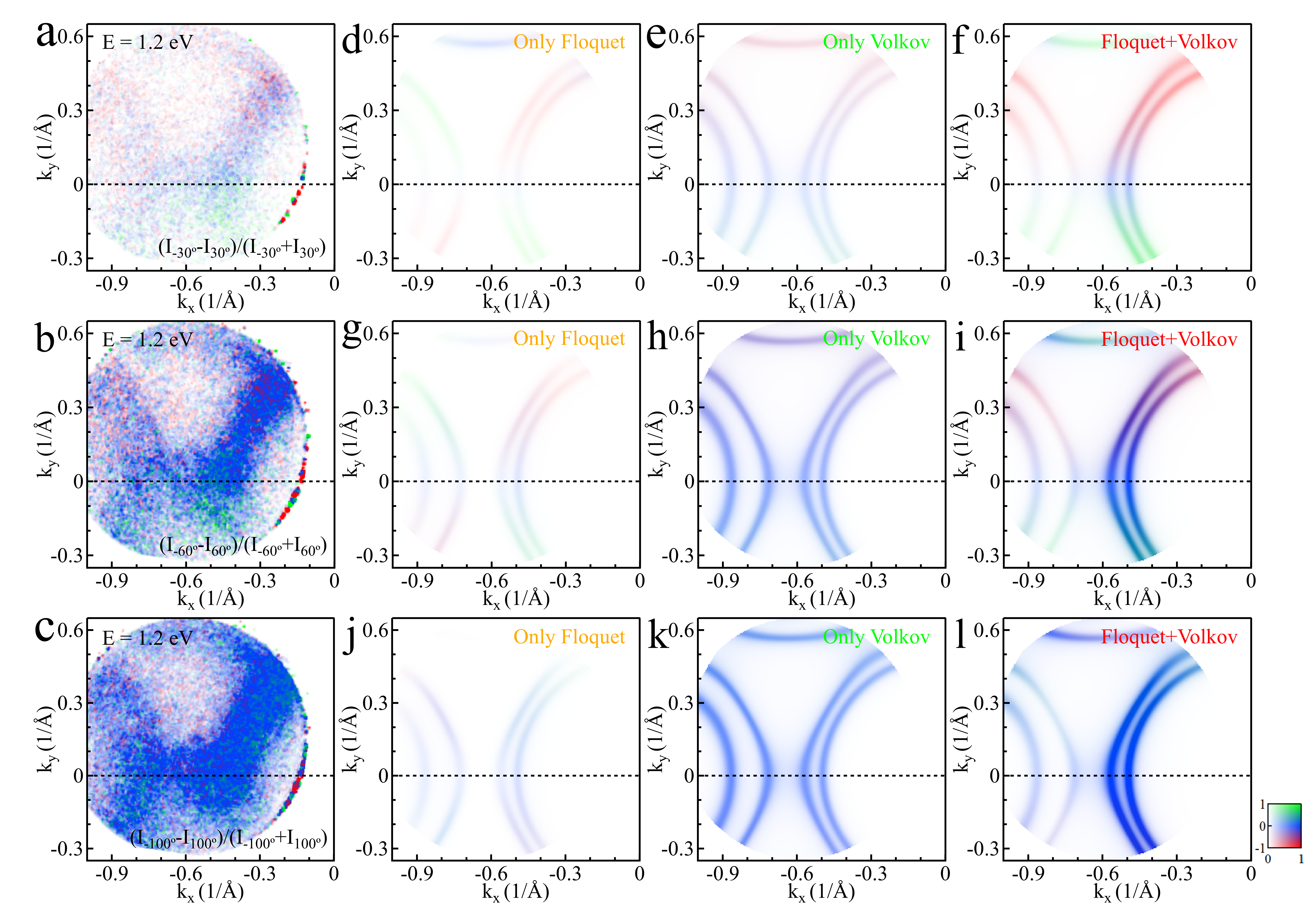}
\end{center}
\caption{\textbf{Polarization-reversal asymmetry and comparison of the Floquet and Volkov channels.}
(a--c) Experimental polarization-reversal asymmetry maps at $E-E_{\mathrm F}=1.2$~eV and $t=0$~fs for $\varphi=30^{\circ}$, $60^{\circ}$ and $100^{\circ}$, respectively, calculated from the opposite-angle intensity pairs in Fig.~3b--g of the main text. The asymmetry is defined in Eq.~\eqref{eq:S_asymmetry}. (d--f) Calculated asymmetry at $\varphi=30^{\circ}$ for the Floquet-only (d), Volkov-only (e) and coherently combined (f) responses. (g--i) Corresponding calculations at $\varphi=60^{\circ}$. (j--l) Corresponding calculations at $\varphi=100^{\circ}$. All panels use the same asymmetry and colour convention.}
\label{fig:S5}
\end{figure*}

The experimental asymmetry maps are constructed from the opposite-angle intensity pairs in Fig.~3b--g of the main text (Supplementary Fig.~\ref{fig:S5}a--c). The $30^{\circ}$ and $60^{\circ}$ pairs exhibit regions of opposite sign, whereas the $100^{\circ}$ pair shows weaker asymmetry over much of the measured region. We apply Eq.~\eqref{eq:S_asymmetry} to the calculated intensity pairs using the same model and global parameter set as in main-text Fig.~4 (Section~9). Neither the Floquet-only nor the Volkov-only response reproduces the complete angular evolution of the momentum-dependent contrast; the coherent response provides the closest overall description (Supplementary Fig.~\ref{fig:S5}d--l). This complementary comparison includes the $\pm100^{\circ}$ pair in addition to the intermediate angles examined in main-text Fig.~4.\\

{\bf 6. Coherent interference versus incoherent addition of Floquet and Volkov contributions}

\begin{figure*}[tbp]
\begin{center}
\includegraphics[width=\textwidth]{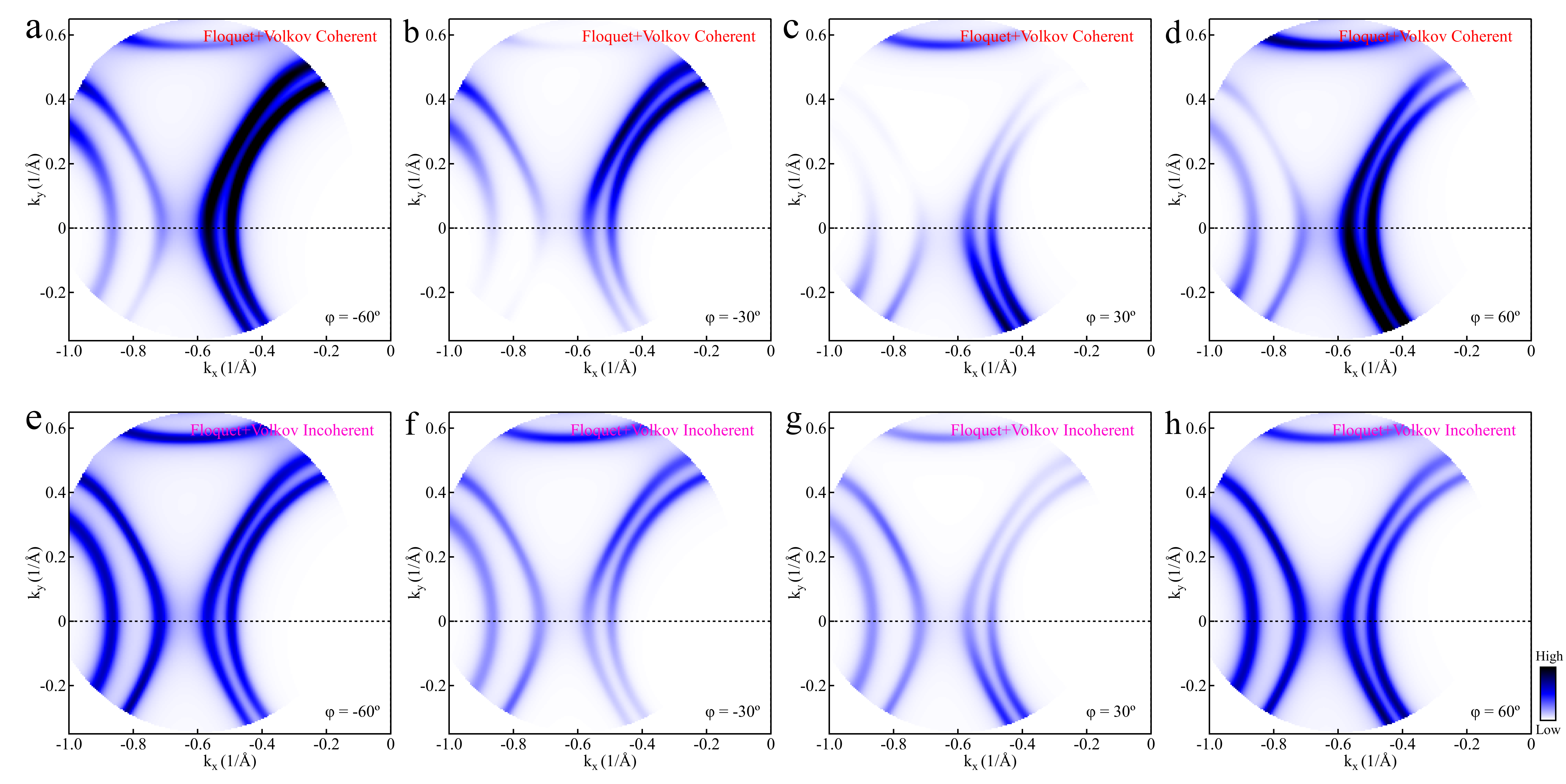}
\end{center}
\caption{\textbf{Coherent interference and incoherent addition of Floquet and Volkov contributions.}
(a--d) Calculated constant-energy sideband maps at $E-E_{\mathrm F}=1.2$~eV for pump-polarization angles $\varphi=-60^{\circ}$, $-30^{\circ}$, $+30^{\circ}$ and $+60^{\circ}$, respectively, including coherent Floquet--Volkov interference. (e--h) Corresponding maps obtained by directly adding the Floquet-only and Volkov-only intensities without interference. Both calculations use the same model parameters (Section~8); the incoherent sum is defined in Eq.~\eqref{eq:S_incoherent_sideband}. The experimental maps at these angles are shown in main-text Fig.~4a,g,m,s, respectively.}
\label{fig:S6}
\end{figure*}

To test whether the branch imbalance in main-text Fig.~4 requires interference rather than the coexistence of the two channels, we compare the coherent response with the direct sum of the Floquet-only and Volkov-only intensities (Supplementary Fig.~\ref{fig:S6}). Both calculations use the same effective fields, band structure and global parameter set; the incoherent sum is defined in Section~9.4.

The incoherent sum retains substantial left-hand spectral weight, particularly at positive polarization angles (Supplementary Fig.~\ref{fig:S6}g,h), and fails to reproduce the right-hand dominance observed in the experimental maps and MDCs of main-text Fig.~4. The coherent response suppresses this excess weight and more closely reproduces the measured branch imbalance (Supplementary Fig.~\ref{fig:S6}a--d). This difference in relative branch intensity cannot be removed by an overall rescaling. Within the common model and parameter set, the comparison therefore supports interference between the initial-state Floquet--Bloch and final-state Volkov amplitudes, rather than a simple addition of their intensities.\\

{\bf 7. Sideband momentum distribution under $s$ polarization}

\begin{figure*}[tbp]
\begin{center}
\includegraphics[width=\textwidth]{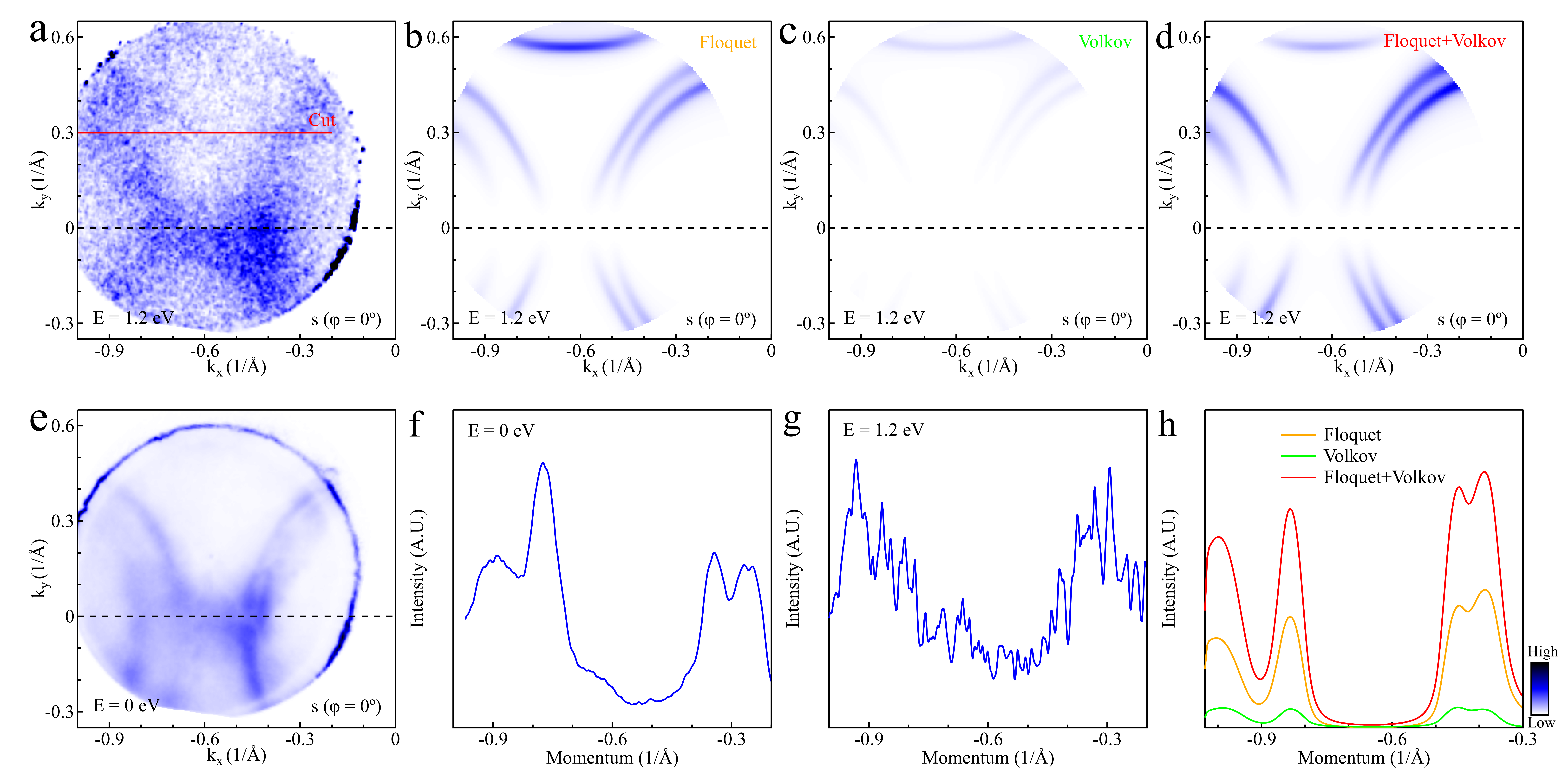}
\end{center}
\caption{\textbf{Comparison of the sideband momentum distribution and calculated responses under $s$ polarization.}
(a) Experimental constant-energy sideband map at $E-E_{\mathrm F}=1.2$~eV and $t=0$~fs under $s$-polarized pumping ($\varphi=0^{\circ}$). The red line marks the horizontal cut at $k_y=0.3$~\AA$^{-1}$. (b--d) Corresponding Floquet-only, Volkov-only and coherent Floquet--Volkov calculations, respectively. (e) Reference Fermi-surface map at $E-E_{\mathrm F}=0$. (f) Experimental momentum distribution curve (MDC) extracted from (e) along the momentum cut defined in (a). (g) Experimental sideband MDC extracted from (a). (h) Calculated sideband MDCs extracted from (b--d) along the same cut: Floquet-only (orange), Volkov-only (green) and coherent Floquet--Volkov (red). The calculations use the global parameter set specified in Section~9.}
\label{fig:S7}
\end{figure*}

At $s$ polarization ($\varphi=0^{\circ}$), the sideband retains the parent-band Fermi-surface contours and has comparable spectral weight on the left- and right-hand sides of the momentum cut (Supplementary Fig.~\ref{fig:S7}a,e--g). The Floquet-only, Volkov-only and coherent calculations all reproduce this approximate balance (Supplementary Fig.~\ref{fig:S7}b--d,h), unlike the pronounced branch imbalance at $\varphi=\pm30^{\circ}$ and $\pm60^{\circ}$ in main-text Fig.~4. The $s$ momentum profile alone therefore does not distinguish the two photoemission pathways.

With the same global parameter set used in the main text (Section~9), the calculated Volkov-only intensity at $s$ is much smaller than the Floquet-only intensity (Supplementary Fig.~\ref{fig:S7}h). This is a model-dependent channel decomposition. The finite measured $s$-polarized yield can also contain a photoexcited background or contributions from imperfect linear polarization, so its magnitude alone does not identify an initial-state response. The branch-resolved comparisons away from $s$ provide the additional constraint. We next examine how the momentum distribution evolves upon small rotations away from $s$.\\

{\bf 8. Polarization dependence near the $s$ configuration}

\begin{figure*}[tbp]
\begin{center}
\includegraphics[width=\textwidth]{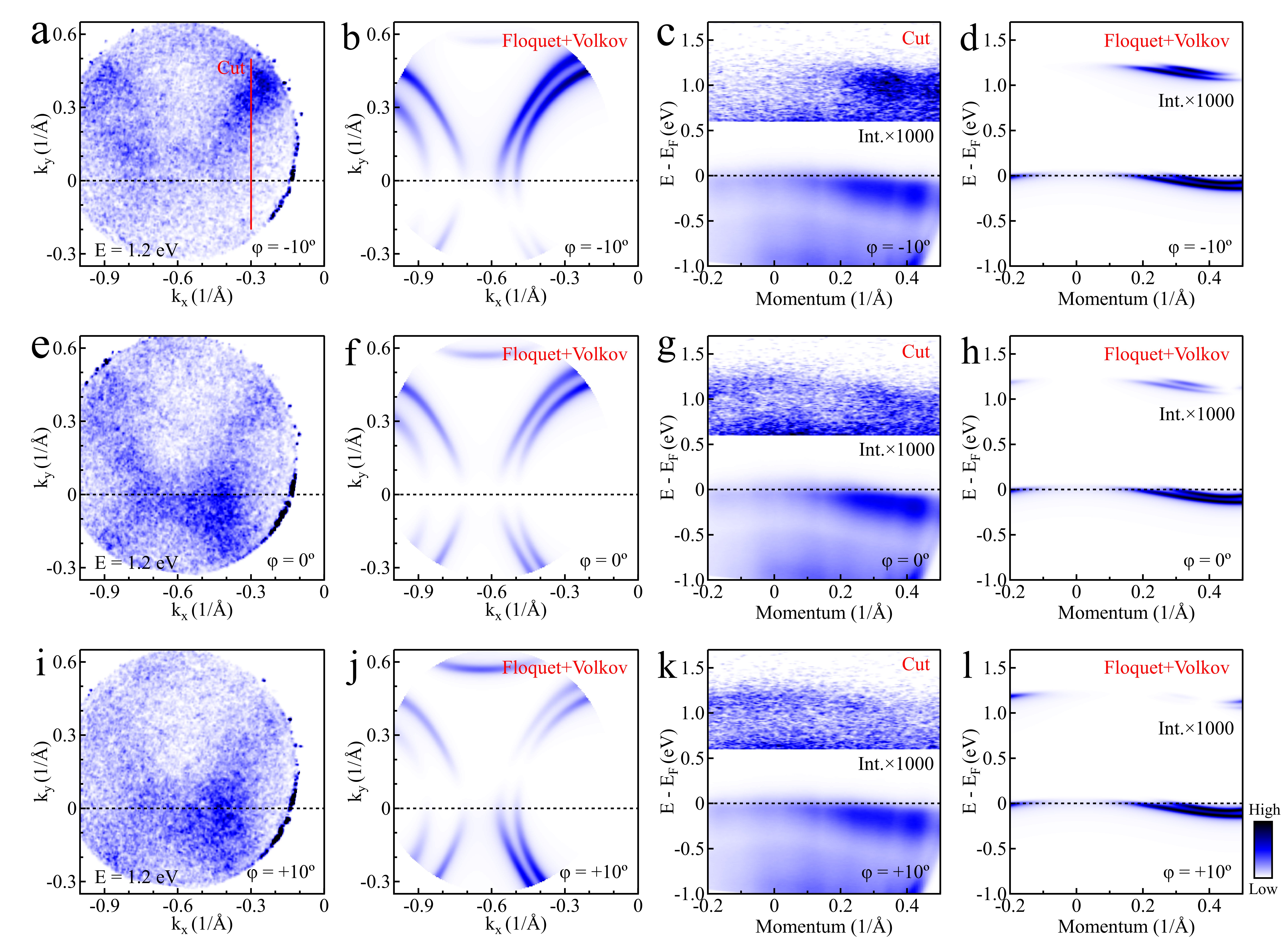}
\end{center}
\caption{\textbf{Comparison of experiment and coherent Floquet--Volkov calculations near $s$ polarization.}
(a--d) Results for $\varphi=-10^{\circ}$ at $t=0$~fs. (a) Experimental constant-energy sideband map at $E-E_{\mathrm F}=1.2$~eV. The red line marks the cut along $k_y$ used for the energy--momentum spectra. (b) Corresponding coherent Floquet--Volkov calculation. (c) Experimental energy--momentum spectrum along the cut in (a). (d) Calculated spectrum along the same cut. (e--h) Same as (a--d), but for $\varphi=0^{\circ}$ ($s$). (i--l) Same as (a--d), but for $\varphi=+10^{\circ}$. The same momentum cut is used at all three angles. Panels (e,f) reproduce Supplementary Fig.~\ref{fig:S7}a,d, respectively, to facilitate comparison across $s$. In (c,d,g,h,k,l), the intensity at $E-E_{\mathrm F}>0.6$~eV is multiplied by 1000 for visibility.}
\label{fig:S8}
\end{figure*}

Small rotations away from $s$ produce distinct changes in the replica weight despite the weak overall sideband yield. The coherent calculation captures the redistribution between $\varphi=-10^{\circ}$ and $+10^{\circ}$, including the finite response at $s$, with the same parameter set used for the larger angles (Supplementary Fig.~\ref{fig:S8}a,b,e,f,i,j).

The corresponding energy--momentum spectra show that this polarization dependence also follows the dispersing replica, whose separation from the parent band remains approximately one pump photon energy (Supplementary Fig.~\ref{fig:S8}c,d,g,h,k,l). Thus, although the balanced $s$ profile alone does not identify the dressing pathway, the evolution on either side of $s$ provides a further test of the coherent response beyond the intermediate-angle comparison in main-text Fig.~4.\\

{\bf 9. Photoemission model and calculation parameters}

\noindent\textbf{9.1. Equilibrium band structure and spectral weight}

We model the sideband measured by time- and angle-resolved photoemission spectroscopy (tr-ARPES) using two Nb-$4d$-derived bands and the coherent sum of the initial-state Floquet--Bloch and final-state Volkov amplitudes. The equilibrium electronic structure around the Fermi level is described by a tight-binding representation fitted to equilibrium angle-resolved photoemission spectroscopy (ARPES) data~\cite{DJRahn_PRB2012_KRossnagel,ASoumyanarayanan_PNAS2013_JEHoffman}. With $\mathbf{k}=(k_x,k_y)$, $\xi=ak_x/2$ and $\eta=\sqrt{3}ak_y/2$, the dispersion of band $b$ is
\begin{equation}
\begin{aligned}
\varepsilon_b(\mathbf{k})={}&t_{0b}
+t_{1b}\left(2\cos\xi\cos\eta+\cos2\xi\right)\\
&+t_{2b}\left(2\cos3\xi\cos\eta+\cos2\eta\right)\\
&+t_{3b}\left(2\cos2\xi\cos2\eta+\cos4\xi\right)\\
&+t_{4b}\left(\cos\xi\cos3\eta+\cos5\xi\cos\eta
+\cos4\xi\cos2\eta\right)\\
&+t_{5b}\left(2\cos3\xi\cos3\eta+\cos6\xi\right),
\end{aligned}
\label{eq:S_tb}
\end{equation}
where $a=3.44$~\AA{} is the in-plane lattice constant and the parameters $t_{0b},\ldots,t_{5b}$ are taken from Ref.~\cite{ASoumyanarayanan_PNAS2013_JEHoffman}. The group velocity is $\mathbf{v}_b(\mathbf{k})=\hbar^{-1}\nabla_{\mathbf{k}}\varepsilon_b(\mathbf{k})$. With energies measured relative to $E_{\mathrm F}$, we represent the equilibrium spectral function by a Lorentzian within a standard spectral-function description of photoemission~\cite{FBoschini_RMP2024_ADamascelli},
\begin{equation}
A_b^{(0)}(\mathbf{k},E)=
\frac{1}{\pi}\frac{\Gamma}
{\left[E-\varepsilon_b(\mathbf{k})\right]^2+\Gamma^2},
\end{equation}
and the corresponding equilibrium photoemission weight is
\begin{equation}
S_b^{(0)}(\mathbf{k},E)=
C_b f_{\sigma_F}(E)A_b^{(0)}(\mathbf{k},E),
\label{eq:S_equilibrium_weight}
\end{equation}
where $f_{\sigma_F}$ is a broadened Fermi function and $C_b$ is an effective probe-photoemission factor. To isolate the polarization dependence generated by Floquet--Volkov phase modulation, $C_b$ is taken to be momentum independent and identical for the two bands.

\noindent\textbf{9.2. Effective fields at the solid--vacuum interface}

The pump-polarization angle is denoted by $\varphi$, with $\varphi=0^\circ$ corresponding to $s$ polarization along $y$ and $\varphi=\pm90^\circ$ to $p$ polarization in the pump's $x$--$z$ incidence plane. The $z$ axis is the surface normal and $x,y$ lie in the (0001) surface. The pump incidence angle is $\theta_i=22^\circ$ from the surface normal. The probe is non-collinear with the pump and its polarization is held fixed. The approximately $23^\circ$ pump--probe separation described in the main-text Methods is not $\theta_i$ and does not replace it in the pump Fresnel coefficients. At the pump photon energy of $1.20$~eV, the model uses the relative dielectric components $\varepsilon_{\parallel}=4.7+3.4i$ and $\varepsilon_{\perp}=3.5+1.1i$~\cite{Munkhbat2022}. The Fresnel reflection coefficients are~\cite{BornWolf}
\begin{equation}
r_s=
\frac{\cos\theta_i-\sqrt{\varepsilon_{\parallel}-\sin^2\theta_i}}
{\cos\theta_i+\sqrt{\varepsilon_{\parallel}-\sin^2\theta_i}},
\end{equation}
and
\begin{equation}
r_p=
\frac{\sqrt{\varepsilon_{\parallel}}\cos\theta_i-
\sqrt{1-\sin^2\theta_i/\varepsilon_{\perp}}}
{\sqrt{\varepsilon_{\parallel}}\cos\theta_i+
\sqrt{1-\sin^2\theta_i/\varepsilon_{\perp}}}.
\end{equation}
Writing the incident electric-field amplitude as $E_0$, the outgoing photoelectron samples the superposition of the incident and reflected fields just outside the surface,
\begin{equation}
\mathbf{E}_{\mathrm{out}}(\varphi)=E_0
\begin{pmatrix}
(1-r_p)\sin\varphi\cos\theta_i\\
(1+r_s)\cos\varphi\\
-(1+r_p)\sin\varphi\sin\theta_i
\end{pmatrix},
\end{equation}
whereas the field acting on the Bloch electrons inside the material follows the electromagnetic boundary conditions,
\begin{equation}
\mathbf{E}_{\mathrm{in}}(\varphi)=E_0
\begin{pmatrix}
(1-r_p)\sin\varphi\cos\theta_i\\
(1+r_s)\cos\varphi\\
-\dfrac{1+r_p}{\varepsilon_{\perp}}\sin\varphi\sin\theta_i
\end{pmatrix}.
\end{equation}
Together, these expressions determine the interfacial field components from the incident pump geometry and the complex dielectric response of the material. The internal and near-surface vacuum fields drive initial-state Floquet and final-state Volkov dressing, respectively. Metallic screening is included through the complex dielectric tensor in the Fresnel calculations.

\noindent\textbf{9.3. Floquet--Bloch and Volkov phase modulation}

We use $e>0$ for the elementary charge and $\Omega=\omega_{\mathrm{pump}}$ for the pump angular frequency. The complex field amplitudes retain the relative phases introduced by the interface.

Initial-state dressing is treated within the intraband Houston approximation~\cite{Houston1940,Krieger1986}. Its dimensionless phase modulation is
\begin{equation}
\beta_b(\mathbf{k}_{\parallel},\varphi)=
\frac{e}{\hbar\Omega^2}
\mathbf{v}_b(\mathbf{k}_{\parallel})\cdot
\mathbf{E}_{\mathrm{in},\parallel}(\varphi).
\label{eq:S_floquet_phase}
\end{equation}
The first-order Floquet sideband then carries the weight $J_1^2(|\beta_b|)$, where $J_1$ is the first-order Bessel function. Final-state dressing is treated within the free-electron Volkov approximation~\cite{STPark_PRA2014_STPark,MKeunecke_PRB2020_SMathias}, with
\begin{equation}
\alpha_{\mathrm V}(\mathbf{k}_{\parallel},E,\varphi)=
C_{\mathrm V}\frac{e}{m_{\mathrm e}\Omega^2}
\mathbf{E}_{\mathrm{out}}(\varphi)\cdot\mathbf{k}_f,
\label{eq:S_volkov_phase}
\end{equation}
where $\mathbf{k}_f=(\mathbf{k}_{\parallel},k_z)$ and
$k_z=\sqrt{2m_{\mathrm e}E_{\mathrm{kin}}/\hbar^2-k_{\parallel}^2}$. Here $E_{\mathrm{kin}}=h\nu_{\mathrm{probe}}-\Phi+E$ is the photoelectron kinetic energy referenced to the sample vacuum level, $h\nu_{\mathrm{probe}}=18$~eV and $\Phi$ is the sample work function. 
The dimensionless scale $C_{\mathrm V}$ multiplies the Volkov phase modulation after the Fresnel fields have been calculated. It accounts phenomenologically for deviations of the actual photoemission final state from the idealized free-electron description; it is not an additional Fresnel coefficient.

\noindent\textbf{9.4. Coherent sideband intensity and incoherent control}

Because the two pathways reach the same detected sideband, their phase modulations combine at the amplitude level. The final state enters the transition amplitude as a bra, giving the total modulation $\gamma_b=\beta_b-\alpha_{\mathrm V}$ and the first-order sideband intensity~\cite{STPark_PRA2014_STPark}
\begin{equation}
I_{+1}^{\mathrm{calc}}(\mathbf{k}_{\parallel},E,\varphi)=
\sum_b S_b^{(0)}(\mathbf{k}_{\parallel},E-\hbar\Omega)
J_1^2\!\left(
\left|\beta_b(\mathbf{k}_{\parallel},\varphi)
-\alpha_{\mathrm V}(\mathbf{k}_{\parallel},E,\varphi)\right|
\right).
\label{eq:S_calculated_sideband}
\end{equation}
In the weak-modulation limit,
\begin{equation}
J_1^2\!\left(|\beta_b-\alpha_{\mathrm V}|\right)
\simeq\frac{1}{4}\left[
|\beta_b|^2+|\alpha_{\mathrm V}|^2
-2\,\operatorname{Re}\!\left(\beta_b\alpha_{\mathrm V}^{*}\right)
\right],
\label{eq:S_weak_interference}
\end{equation}
which explicitly shows the interference term. Floquet-only and Volkov-only distributions are obtained by setting $\alpha_{\mathrm V}=0$ and $\beta_b=0$, respectively; both amplitudes are retained in the coherent calculation. The same procedure and parameter set are used for every pump-polarization angle.

For the incoherent control in Supplementary Fig.~\ref{fig:S6}, we add the separately calculated Floquet-only and Volkov-only intensities,
\begin{equation}
\begin{aligned}
I_{+1}^{\mathrm{inc}}(\mathbf{k}_{\parallel},E,\varphi)
&=I_{+1}^{\mathrm F}(\mathbf{k}_{\parallel},E,\varphi)
+I_{+1}^{\mathrm V}(\mathbf{k}_{\parallel},E,\varphi)\\
&=\sum_b S_b^{(0)}(\mathbf{k}_{\parallel},E-\hbar\Omega)
\left[J_1^2(|\beta_b|)+J_1^2(|\alpha_{\mathrm V}|)\right].
\end{aligned}
\label{eq:S_incoherent_sideband}
\end{equation}
The arguments of $\beta_b$ and $\alpha_{\mathrm V}$ are the same as in Eq.~\eqref{eq:S_calculated_sideband}. This direct intensity sum contains no interference between the two channels. In the weak-modulation limit, it reduces to the first two terms in Eq.~\eqref{eq:S_weak_interference}, without the cross term. The parameters are kept identical to those of the coherent calculation; no additional channel weights are introduced.

\noindent\textbf{9.5. Parameters and comparison with experiment}

The experimental parameters are $\hbar\Omega=1.20$~eV and $\theta_i=22^{\circ}$. The relative amplitudes and phases of the effective field components are determined by the dielectric tensor and incidence geometry. The relative Volkov scale is calibrated to the measured $p$-to-$s$ integrated intensity ratio of 18.6 in Fig.~3h of the main text after the normalization described in Section~4, giving $C_{\mathrm V}\approx0.15$. With the other parameters fixed, this value reduces the unscaled free-electron Volkov coupling to reproduce the measured contrast. The Fermi-edge broadening is set to $\sigma_F=15$~meV, comparable to the experimental energy resolution of the 18~eV setup, and the Lorentzian half-width $\Gamma=18$~meV is used as an effective band linewidth.

The constant-energy calculations are evaluated at $E-E_{\mathrm F}=1.2$~eV. The calculated pairs at $\pm\varphi$ are processed with Eq.~\eqref{eq:S_asymmetry} to obtain Supplementary Fig.~\ref{fig:S5}. The coherent and incoherent intensity maps in Supplementary Fig.~\ref{fig:S6} are evaluated using Eqs.~\eqref{eq:S_calculated_sideband} and~\eqref{eq:S_incoherent_sideband}, respectively; retaining the energy dependence gives the spectra in Supplementary Fig.~\ref{fig:S8}. The dielectric response, linewidths and Volkov scale are held fixed throughout these comparisons. The isolated-channel curves are therefore common-parameter controls, rather than independently optimized fits.

\noindent\textbf{9.6. Momentum distribution curves and comparison with experiment}

The MDCs in main-text Fig.~4 are extracted from the constant-energy sideband maps along the horizontal cut marked in Fig.~4a. The same cut is used at $\varphi=-60^{\circ}$, $-30^{\circ}$, $+30^{\circ}$ and $+60^{\circ}$ for the experiment and all three calculated responses. Writing $I_{\varphi}^{(j)}(k_x,k_y,E)$ for the corresponding intensity distribution, with $E$ measured relative to $E_{\mathrm F}$, the fixed-energy profile is
\begin{equation}
M_{\varphi}^{(j)}(k_x)=
I_{\varphi}^{(j)}(k_x,k_y^{\mathrm{cut}},E_{\mathrm{cut}}),
\qquad E_{\mathrm{cut}}=1.2~\mathrm{eV},
\label{eq:S_mdc}
\end{equation}
where $k_y^{\mathrm{cut}}$ is set by the line in main-text Fig.~4a and $j$ denotes the experiment, Floquet-only, Volkov-only or coherent Floquet--Volkov response. The experimental MDCs are shown in main-text Fig.~4e,k,q,w; the three calculated MDCs are compared in Fig.~4f,l,r,x. The $s$ sideband MDCs in Supplementary Fig.~\ref{fig:S7}g,h are extracted analogously along the cut in Supplementary Fig.~\ref{fig:S7}a. The model parameters are those specified in Section~9.5.

The experimental profiles retain stronger weight on the right-hand side of the cut at all four angles. The Floquet-only response favours the left-hand side at positive angles, whereas the Volkov-only response overestimates the relative weight of the left-hand features. The coherent response gives the closest overall description of this branch-intensity imbalance across the four angles, consistent with the full momentum maps and the asymmetry analysis in Section~5. The comparison concerns the relative momentum-dependent spectral weight; it does not rely on absolute intensity calibration between experiment and calculation.

\noindent\textbf{9.7. Sensitivity to the relative Volkov scale}

\begin{figure*}[tbp]
\begin{center}
\includegraphics[width=\textwidth]{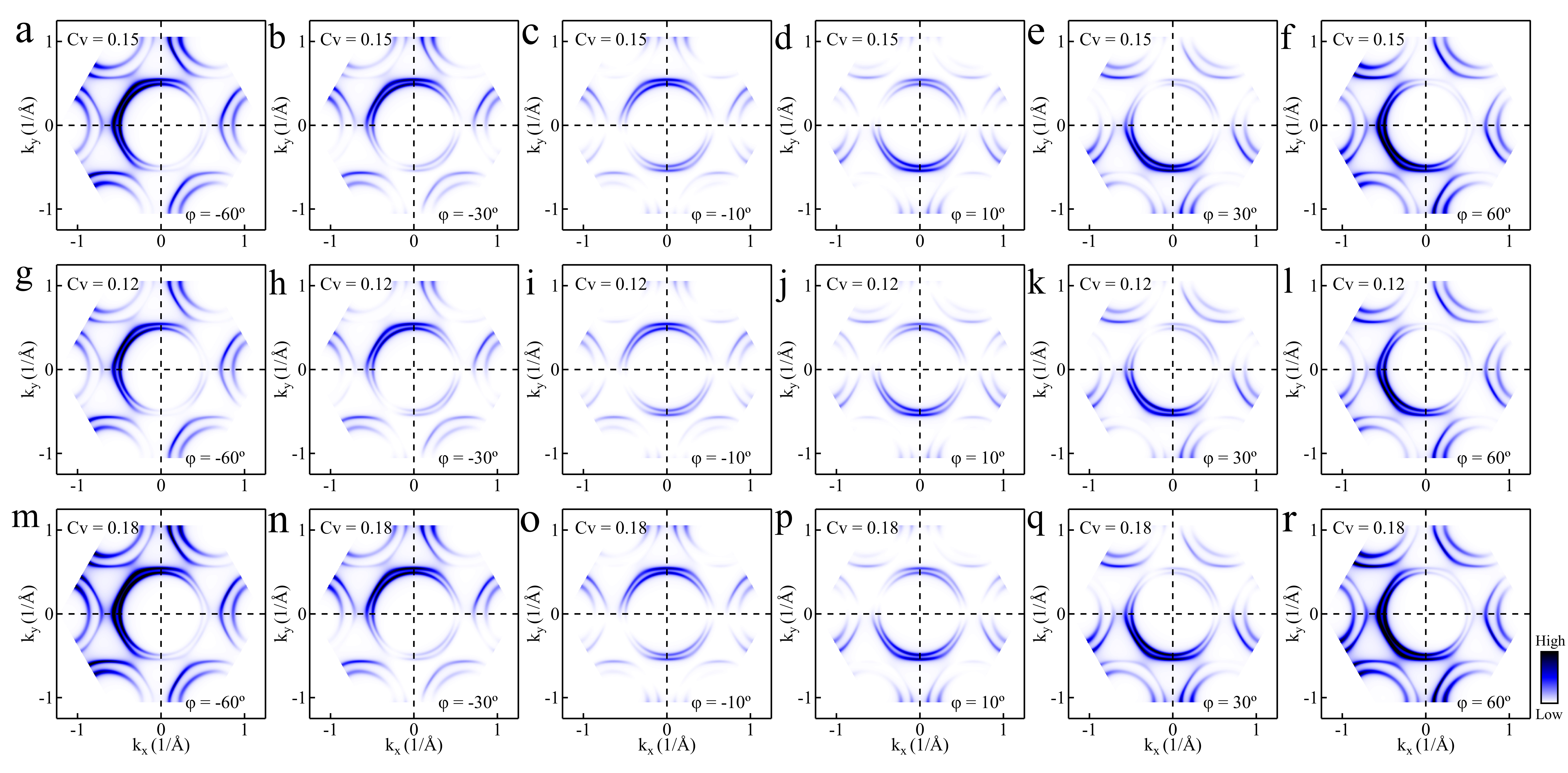}
\end{center}
\caption{\textbf{Sensitivity of the coherent Floquet--Volkov momentum texture to the relative Volkov scale.}
(a--f) Calculated constant-energy sideband maps at $E-E_{\mathrm F}=1.2$~eV for pump-polarization angles $\varphi=-60^{\circ}$, $-30^{\circ}$, $-10^{\circ}$, $+10^{\circ}$, $+30^{\circ}$ and $+60^{\circ}$, respectively, using the calibrated value $C_{\mathrm V}=0.15$. (g--l) Corresponding maps with $C_{\mathrm V}$ reduced by 20\%. (m--r) Corresponding maps with $C_{\mathrm V}$ increased by 20\%. The values displayed in the panels, 0.15, 0.12 and 0.18, are rounded. All other model parameters are held fixed. The principal polarization-dependent features persist across this range (Section~9.7).}
\label{fig:S9}
\end{figure*}

To test whether the calculated momentum texture depends sensitively on the calibrated value of $C_{\mathrm V}$, we vary this parameter by $\pm20\%$ while keeping all other parameters fixed (Supplementary Fig.~\ref{fig:S9}). This variation is a sensitivity test, not a statistical confidence interval. The polarization-dependent redistribution of sideband weight and the dominant contour features persist across the tested range, including the small-angle evolution on either side of $s$. 
The $p$-to-$s$ integrated intensity ratio changes from 14.5 at the reduced scale to 18.6 at the calibrated scale and 22.4 at the increased scale. Thus, the integrated contrast constrains $C_{\mathrm V}$ quantitatively, while the momentum texture remains qualitatively stable under these variations. We retain $C_{\mathrm V}=0.15$ for the comparisons with experiment.

\vspace{3mm}

\newpage

